\documentclass[a4paper,fleqn,usenatbib]{mnras}

\usepackage{newtxtext,newtxmath}

\usepackage[T1]{fontenc}
\usepackage{ae,aecompl}
\usepackage{tgtermes}

\usepackage{graphicx}	
\usepackage{amsmath}	
\usepackage{amssymb}	
\usepackage{hyperref}

\def  \mxb {MXB 1659-298}
\def  \mxbs {MXB 1659-298 }

\def  \axs {AX J1745.6-2901 }
\def \exo {EXO 0748-676}
\def \exos {EXO 0748-676 }

\title[The new ephemeris of \mxb]{A possible solution of the puzzling 
  variation of the orbital period  of \mxb }

\author[R. Iaria et al.]{
R. Iaria,$^{1}$\thanks{E-mail: rosario.iaria@unipa.it}
A. F. Gambino,$^{1}$
T. Di Salvo,$^{1}$
L. Burderi,$^{2}$
M. Matranga,$^{1}$ \newauthor
A. Riggio,$^{2}$
A. Sanna,$^{2}$
F. Scarano,$^{2}$
and   A. D'A\`i$^{3}$\\
$^{1}$Dipartimento di Fisica e Chimica, Universit\`a di Palermo, via Archirafi 36 - 90123 Palermo, Italy\\
$^{2}$Dipartimento di Fisica, Universit\`a degli Studi di Cagliari, SP Monserrato-Sestu, KM 0.7, Monserrato, 09042 Italy\\
$^{3}$INAF/IASF Palermo, via Ugo La Malfa 153, I-90146 Palermo, Italy.}

\date{Accepted XXX. Received YYY; in original form ZZZ}

\pubyear{2016}

\begin{document}
\label{firstpage}
\pagerange{\pageref{firstpage}--\pageref{lastpage}}
\maketitle

\begin{abstract}
  \mxbs is a transient neutron star Low-Mass X-ray binary system that
  shows eclipses with a periodicity of 7.1 hr.  \mxbs went to outburst
  in August 2015 after 14 years of quiescence. We investigate the
  orbital properties of this source with a baseline of 40 years
  obtained combining the eight eclipse arrival times present in
  literature with 51 eclipse arrival times collected during the last
  two outbursts.  A quadratic ephemeris does not fit the delays
  associated with the eclipse arrival times and the addition of a
  sinusoidal term with a period of $2.31 \pm 0.02$ yr is required. 
 We infer a binary orbital period of
  $P=7.1161099(3)$ hr and an orbital period derivative of
  $\dot{P}=-8.5(1.2) \times 10^{-12}$ s s$^{-1}$.  We show that the
  large orbital period derivative can be explained with a highly non
  conservative mass transfer scenario in which more than 98\% of the
  mass provided by the companion star leaves the binary system. We
  predict an orbital period derivative value of
  $\dot{P}=-6(3) \times 10^{-12}$ s s$^{-1}$ and constrain the
  companion star mass between $\sim$0.3 and $ 0.9 \pm 0.3$
  M$_{\odot}$. Assuming that the companion star is in thermal
  equilibrium the periodic modulation can be due to either   a
  gravitational quadrupole coupling due to variations of the
  oblateness of the companion star or with the presence of a third
  body of mass M$_3 >21 $ Jovian masses.
\end{abstract}




\section{Introduction}

One of the most direct evidence for binary orbital motion is the
presence of eclipse of the central source by a companion star.  For
Low Mass X-ray Binaries (LMXBs) with inclination angles between
75$^{\circ}$ and 80$^{\circ}$ the X-ray emission may be totally
shielded by the companion star.  As the companion transits between the
X-ray central source and the observer the light curves show total
eclipses.  For inclination angles between 80$^{\circ}$ and
  90$^{\circ}$ the LMXB is observed as an Accretion Disc Corona (ADC)
  source. In this case the observed X-ray emission comes from an
  extended corona that can reach the outer region of the accretion disc. The light curves of the ADC sources show an almost
  sinusoidal modulation and partial eclipses.  The modulation of the
  light curve is generally explained with the presence of a
  geometrically thick disc whose height varies depending on the
  azimuthal angle and occults part of the X-ray emission. Since the companion star
does not shield the whole extended corona the observed eclipses are partial; the
  prototype of the ADC sources is X1822-371 \citep[see
  e.g.][and references therein]{Iaria_11,Iaria_13,Iaria_15b}.

 Total eclipses represent a good time reference, which is ideal to 
perform timing analysis of the binary orbital period, e.g.
the O-C
method is usually applied to refine the orbital period or trace
orbital period changes \citep[see][for a recent review]{Chou_14}.
To date, 12 LMXBs show total eclipses in their light curve.
One of the best studied eclipsing X-ray source is \exo, as it was active
 for more than 20 years \citep[see][and
references therein]{Wolff_09}.

 The eclipsing LMXB \mxbs was discovered by \cite{Lewin_76} in
  1976. The light curve showed type-I X-ray bursts, 
thus revealing that the compact object was an accreting neutron star.
The source was observed in outburst up to 1978 with {\it SAS3}
and {\it HEAO} \citep[][]{Com_83,Com_84,Com_89}.   Eclipses were
  firstly 
reported by
by \cite{Com_84}, which estimated a periodicity of 7.1 hr.
\cite{Com_89} analysed two whole eclipses estimating two eclipse
arrival times. From 1978 up to 1999 the region  containing 
\mxbs was monitored by the X-ray observatories onboard {\it Hakucho},
{\it EXOSAT} and {\it ROSAT}, but the source was never detected
\cite[see][]{Com_89, Verbunt_01}. On April 1999 the Wide Field Cameras
onboard {\it BeppoSAX} observed the source in outburst again
\citep{Zand_99}. This new outburst continued up to September
2001. During the outburst \mxbs was observed with the Proportional
Counter Array (PCA) onboard {\it Rossi X-ray Timing Explorer}
\citep[{\it RXTE}, see e.g.][]{Wachter_00}, with the Narrow Field
 Instruments (NFI) onboard
{\it BeppoSAX} \citep{Ost_01} and with {\it XMM-Newton}.  From the
analysis of the RXTE light curves of the source, \cite{Wachter_00}
obtained four eclipse arrival times and found an orbital period
derivative of $(-7.2 \pm 1.8) \times 10^{-11}$ s s$^{-1}$
suggesting that the orbit of the binary system is
shrinking. \cite{Ost_01} obtained two eclipse arrival times from a
{\it BeppoSAX}/NFI observation and combining their data with those
present in literature found that the orbital period derivative,
$\dot{P}_{\rm orb}$, is positive with a value of 
$ (7.4 \pm 2.0) \times 10^{-12}$ s/s.  \mxbs 
  turned  again on
outburst on 2015 August 21 \citep{Negoro_15} and 
up to  2017 March is still X-ray bright.
 Using data of the X-ray Telescope (XRT) onboard {\it Swift},
\cite{Bahram_16} observed that the unabsorbed flux in the 0.5-10 keV
energy range was $1.5 \times 10^{-10}$, $4.6 \times 10^{-10}$ and
$2.2 \times 10^{-10}$ erg cm$^{-2}$ s$^{-1}$ on 2016 January 28,
February 2 and 11, respectively.
 \begin{figure*}
	\includegraphics[scale=.34,angle=-90]{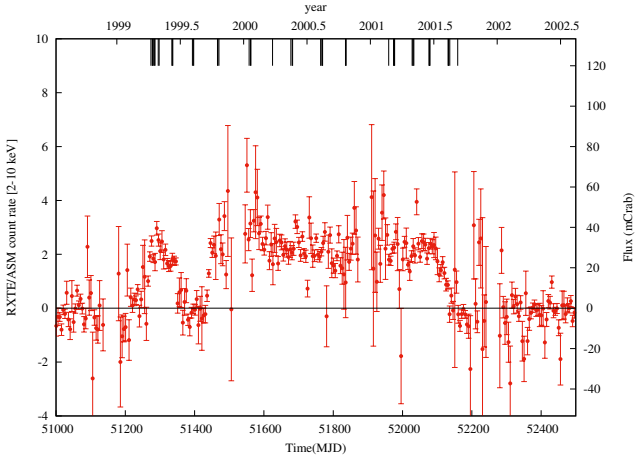}
	\includegraphics[scale=.34,angle=-90]{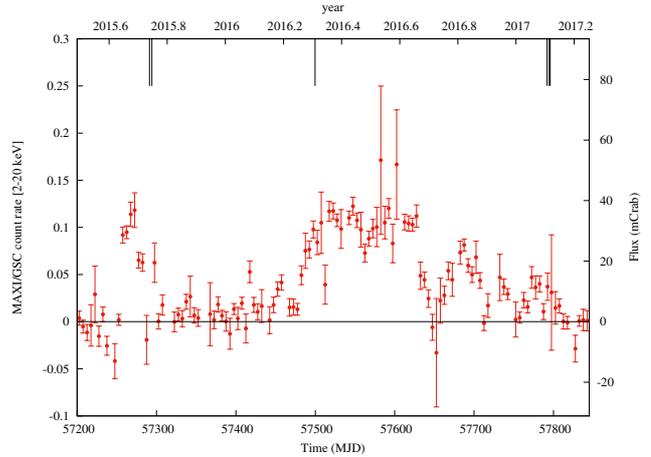}
  \caption{Light curve of \mxbs during the outburst occurred between
    1999 and 2001
    (left panel) and  the latest  started 2015 (right panel). The left panel shows the
    RXTE/ASM light curve in the 2-10 keV, the right panel shows the
    MAXI/GSC light curve in the 2-20 keV energy range; the bin time is
     {\bf five days} for both the light curves. The eclipse arrival times are also indicated.}
    \label{fig:outburst}
\end{figure*} 

\cite{Com_89} measured an eclipse duration, $\Delta T_{\rm ecl}$, of
  $932 \pm 13$ s and an ingress/egress 
  duration of $\Delta T_{\rm ing} = 41 \pm 13$ s and
  $\Delta T_{\rm egr}= 19 \pm 13$ s, respectively. They showed that, 
if the companion star is  a main-sequence star of 0.9
  M$_{\odot}$   with a  temperature close to 5000 K, 
the scale height of the stellar atmosphere should be around 200 km,
corresponding to an ingress/egress duration close to 0.5 s.
 The authors concluded that 
the small value of the  scale height cannot justify the large value of
the measured ingress/egress durations.  Furthermore, \cite{Com_89}
suggested that the observed asymmetry between the ingress and egress
duration could be caused by a one-sided extended corona of size $5 \times
10^5$ km. 
 
From the analysis of four
  eclipses obtained with {\it RXTE}/PCA, \cite{Wachter_00} estimated an
 average eclipse duration of $901.9 \pm 0.8 $ s and average values of
  ingress/egress durations  of $\Delta T_{\rm ing} = 9.1 \pm 3.0$ s and
  $\Delta T_{\rm egr} = 9.5 \pm 3.3$ s.  The authors  proposed  that the large
  spread of  values associated with the ingress/egress times could
  be caused either by flaring activity of the companion star or by the
  presence of an evaporating wind from the surface of the
  companion star  created by irradiation from the X-ray source. 

 \cite{Com_84} discussed the nature of the optical counterpart of
  \mxb, named V2134 Oph, assuming an orbital period of 7.1 hr and an
  eclipse duration of 900 s. They constrained the mass of the
  companion star to be between 0.3 M$_\odot$ and 0.9 M$_\odot$ for an
  inclination angle of the binary system of 90$^\circ$ and
  71$^\circ$\!.5, respectively.  \cite{warner_95} inferred that the
  companion star mass is between 0.75 and 0.78 M$_\odot$ if the
  companion fills its Roche lobe. This range of masses suggests that
  the companion is a K0 main-sequence star.  During the quiescence of
  \mxb, \cite{Wachter_00} measured a magnitude in the $I$-band of
  $ 22.1 \pm 0.3$ mag and \cite{Filippenko_99} measured a magnitude in
  the $R$-band of $23.6 \pm 0.4$ mag.  \cite{Wachter_00} found
  that the value of $(R-I)_0$ is compatible with an early K spectral
  type. Moreover, they suggested that, for a companion star belonging to
  the K0 class, the visual magnitude should be $V=23.6$ mag, value
  that is compatible with the measured lower limit of $V> 23$ mag.

\cite{Gallo_08}, analysing the type-I X-ray bursts observed with
  {\it RXTE}/PCA,  inferred a distance to the source of 
   $9 \pm 2$ and $12 \pm 3$ kpc for a hydrogen-rich and helium-rich
  companion star, respectively. Furthermore, \cite{Win_01} detected
  nearly coherent oscillations with a frequency around 567 Hz during
  type-I X-ray bursts suggesting that the neutron star  could be an
  X-ray millisecond pulsar with a spin period of 1.8 ms.

The interstellar hydrogen column density, $N_{\rm H}$, was estimated by
\cite{Cack_08} during the X-ray quiescence of \mxb. 
Combining {\it Chandra} and {\it XMM-Newton } 
 observations collected between 2001 and 2008 they fitted the
  X-ray spectrum
obtaining $N_{\rm H} = (2.0 \pm 0.1)
\times 10^{21}$
cm$^{-2}$.  Two more recents  {\it Chandra} observations of the source,
taken in 2012 \cite{Cack_13},  seem to suggest an
increase of the  interstellar hydrogen column
density at the value of $(4.7 \pm 1.3) \times 10^{21}$
cm$^{-2}$. The authors
proposed three different scenarios
to explain the increase of $N_{\rm H}$: a) material is building up in the
outer region of the accretion disc, b) the presence of a precessing accretion
disc, and c) sporadic variability during   quiescence due to
low-level accretion. 

Studying the {\it XMM}-Newton spectrum of \mxb, \cite{Sidoli_01}
detected two absorption lines at 6.64 and 6.90
keV associated with the presence of highly ionised iron
 (\ion{Fe}{xxv} and \ion{Fe}{xxvi}
ions) as well as absorption lines associated with highly ionised oxygen and
neon (\ion{O}{viii} 1s-2p,  \ion{O}{viii} 1s-3p, \ion{O}{viii} 1s-4p
and \ion{Ne}{ix} 1s-2p transition) at 0.65, 0.77, 0.81 and 1.0 keV.

In this paper we report the updated ephemeris of \mxbs  combining 45 eclipse
arrival times obtained with {\it XMM-Newton} and {\it RXTE} during the
outburst between 1999 and 2001 and six eclipse arrival times obtained
with {\it XMM-Newton},  {\it NuSTAR} and {\it Swift/XRT} during the
outburst started in 2015. The available temporal baseline allows to
partially constrain the bizarre behaviour of the eclipse arrival
times.

\section{Observations}

During the outburst occurred from 1999 to 2001,  \mxbs was observed  with  {\it
  XMM-Newton}  \citep{Jansen_01} two times: 
on March 22 2000 and on Feb. 20 2001.
The latter observation (obsid. 0008620701) was analysed by
\cite{Sidoli_01} and \cite{Diaz_06}, which studied the spectral
properties of the source during the persistent emission, the dip and
the eclipse, while the former observation  (obsid. 0008620601) was
  never analysed.  During the 2015 outburst, \mxbs was observed with
{\it XMM-Newton} on September 26, 2015.

The European Photon Imaging Camera
\citep[Epic-pn,][]{struder_01} 
onboard {\it XMM-Newton} collected data from the source in timing
mode, with  
exposure times of  10, 32 and 34 ks, respectively.  The Epic-pn light
curve of the observation taken in 2001 shows two eclipses in the light
curve \citep[see Fig. 1 in][]{Sidoli_01}.  To verify the presence
of eclipses in the Epic-pn light curves of the observations taken in
2000 and 2015 we filtered the source events with the Science Analysis System (SAS)
ver. 15.0.0. We reprocessed the Epic-pn events and applied the 
  solar-system barycentre corrections adopting as coordinates
 RA$=255^{\circ}\!.527250$ and Dec$=-29^{\circ}\!.945583$
\cite[see][]{Wij_03}.  During the observation taken in 2000, the light
curve of \mxbs shows an eclipse with a duration of 900 s 
approximately 1400 s after the start time.
 The count rate is 32 c s$^{-1}$ and 1.4 c s$^{-1}$
outside and during the eclipse, respectively. During the
observation taken in 2015, the light curve shows the presence of a
type-I X-ray burst at 12 ks  after the start of the observation. 
The count rate varies
from 32 c s$^{-1}$ at the beginning of the burst up to 320 c s$^{-1}$
at the peak. An intense dipping activity is present at about 20 ks
from the beginning of the observation, a complete eclipse is observed at 26
ks from the start time and an eclipse without the ingress is observed
at beginning of the observation. The count rate out and during the
eclipse is 32 and 1.4 c s$^{-1}$, respectively.

The PCA instrument onboard {\it
  RXTE} \citep{Jahoda_96} observed several
times the source from 1999 to 2001. In our analysis we selected 43
{\it RXTE}/PCA observations  showing the eclipse  and for
  which it is
possible to estimate the ingress and egress time accurately. 
To estimate the eclipse arrival times from the {\it RXTE}/PCA
observations we analysed the standard product  background-subtracted
light curves with a bin time of 0.125 s and  we applied the solar-system barycentric
correction to the events using the ftool {\tt faxbary}. 

{\it NuSTAR} \citep[][]{Harrison_13} observed \mxbs  two times in 2015 and 2016
with both the  independent solid state
photon counting detector modules (FPMA and  FPMB),  with elapsed times 
 of 96 ks and 50 ks, respectively. 
We processed the raw (Level 1) data with the ftool {\tt nupipeline}
(Heasoft ver. 6.19),  obtaining cleaned and calibrated event data (Level 2).
The  solar-system barycentric corrected events of the FPMA and FPMB telescopes
have been obtained  applying the tool {\tt nuproducts} on the Level 2
data. The corresponding light curves 
were  created  selecting a circular extraction region for the source
events with a radius of $49\arcsec\!$ and using the 1.6-20 keV energy
range.
The persistent emission has a count rate of  2 c s$^{-1}$. 
A complete eclipse and an eclipse without the ingress
are observed at 24 and 76 ks from the start time. 
The count rate during the eclipse is 0.02 c s$^{-1}$.  It is also evident the
presence of the  ingress to the eclipse at 49.7 ks from the start time. 
During the second
observation \mxbs is brighter, with a persistent count rate of 20  c
s$^{-1}$, a whole eclipse is
observed   30 ks after the start time of the observation. 
To increase the statistics of the {\it NuSTAR} light curve we summed
the FPMA and FPMB light curve using the ftool {\tt lcmath}. 

During the 2015 outburst, \mxbs was observed several times with {\it
  Swift}/XRT \citep{Gerels_04, Burrows_05}, although only three
observations show a complete eclipse. We obtained further {\it
  Swift}/XRT data as target of opportunity observations performed on
February 8, 10 and 11, 2017 (obsid 0003400266, 0003400267 and
0003400268). All of the three observations cover the whole eclipse.
The XRT data were processed with standard procedures (xrtpipeline
v0.13.1), and with standard filtering and screening criteria with
ftools. For our timing analysis, we also converted the event arrival
times to the solar-system barycentre with the tool {\tt barycorr} and
subtracted the background using the ftool {\tt lcmath}.

The All Sky monitor \citep[ASM,][]{Levine_96}  onboard {\it RXTE} 
monitored the 1999-2001 outburst
(Fig. \ref{fig:outburst}, left panel).  The two {\it XMM-Newton }
observations were performed at a similar ASM count rate of 2.5 c
s$^{-1}$ (about 30 mCrab in flux),  corresponding to the source 
maximum flux.  The outburst showed
a sort of precursor lasting 100 d,  afterwards the flux
decreased up to a value compatible with zero for 86 d, and finally
increased again rapidly reaching a constant flux of 30 mCrab for 700 d.

The Gas Slit Camera \citep[GSC,][]{Mihara_11} onboard the Monitor of All-sky X-ray Image
\citep[MAXI,][]{Matsuoka_09} observed the recent outburst (see Fig. \ref{fig:outburst},
right panel). The morphology of the outburst is similar to the
previous one with a sort of precursor lasting 50 d, a new quiescent
stage lasting 150 d and, after that, an increase of the flux at 30 mCrab lasting 150
d.  The maximum GSC count rate is 0.12 c s$^{-1}$. {\it XMM-Newton}
and {\it NuSTAR} (obsid. 90101013002) observed the source when the GSC
count rate was 0.05 c s$^{-1}$; {\it NuSTAR} observed the source a
second time when \mxbs was brighter with a corresponding GSC count
rate of 0.1 c s$^{-1}$.

\section{Method and Analysis}
\begin{figure}
	\includegraphics[width=\columnwidth]{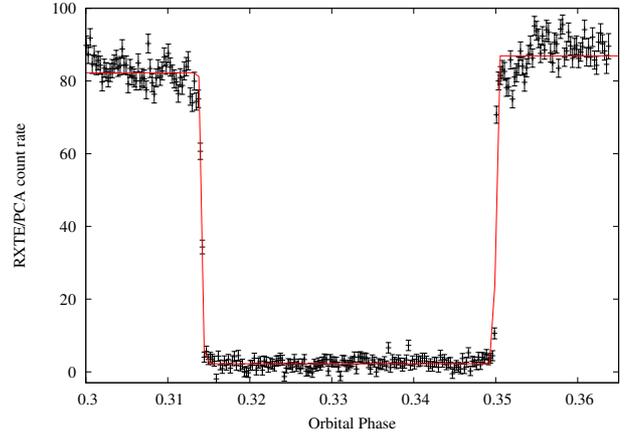}
  \caption{Eclipse of MXB 1659-628 observed by the RXTE/PCA
    instrument (observation
    P40050-04-16-00). The superimposed red function is the
    step-and-ramp function adopted to estimate the eclipse arrival time.}
    \label{eclipse}
\end{figure}
 \begin{figure}
	\includegraphics[width=\columnwidth]{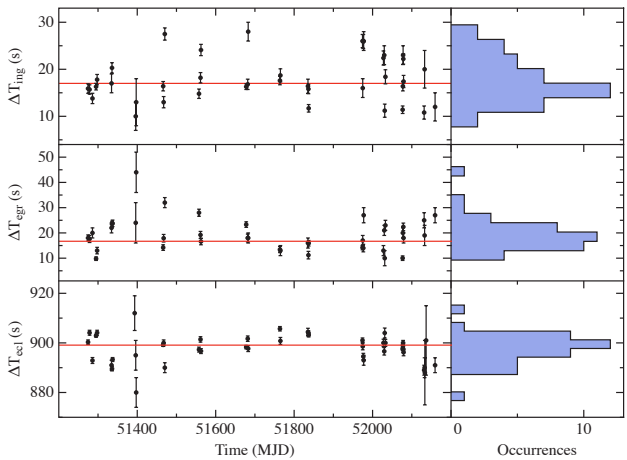}
  \caption{From the top-left to the bottom-left  the ingress, egress and eclipse
    duration, respectively, as function of time.  The  values are
    obtained from the RXTE/PCA eclipses analysed  in this   work.  The red lines indicate the average values for each
    duration. From the top-right to the bottom-right we  show the
    histograms  of the occurrences of the ingress, egress and eclipse
    duration.   }
    \label{duration}
\end{figure} 
\begin{table*}
	\centering
	\caption{Journal of the X-ray eclipse arrival times of \mxb}
	\label{tab:table_delay}
	\begin{tabular}{l l r r  c  l l r r c } 
		\hline
Point&
Eclipse Time&
Cycle&
Delay&
Ref.   &
Point&
Eclipse Time&
Cycle&
Delay&
Ref.   \\

 &
(MJD;TDB)&
 &
(s)& 
    &
&
(MJD;TDB)&
 &
(s)& 
    \\
		\hline
1 &
43$\;$058.7260(2)&
0 &
0(13)&
[1],[2] &
31&
51$\;$769.43726(2)&
29378 &
107.0(1.3)&
[4]\\

2 & 
43$\;$574.6441(2) &
1740 &
26(13)&
[1],[2]&
32&
51$\;$835.261275(9)&
29600&
106.9(7) &
[3]\\

3&
51$\;$273.978079(2)&
27707&
96.46(13)&
[2]&
33&
51$\;$836.447292(6)&
29604&
106.8(5) &
[3]\\

4&
51$\;$274.571102(8)&
27709&
97.7(7)&
[3]&
34&
51$\;$837.040274(5)&
29606&
104.4(4) &
[3]\\

5&
51$\;$277.832626(4)&
27720 &
95.4(3)&
[2]&
35&
51$\;$960.08961(2)&
30021&
101(2)&
[3]\\

6&
51$\;$278.425648(10) &
27722 &
96.5(9)&
[3]&
36&
51$\;$974.321836(6)&
30069&
101.4(5) &
[3]\\

7&
51$\;$281.687174(4)&
27733&
94.5(3)&
[2]&
37&
51$\;$974.914836(8)&
30071&
100.6(7)&
[3]\\

8&
51$\;$283.762726(3)&
27740 &
96.2(3)&
[2]&
38&
51$\;$976.397381(8)&
30076&
102.5(6) &
[3]\\

9&
51$\;$285.838220(11)&
27747 &
93.0(9)&
[3]&
39&
51$\;$977.286855(12)&
30079&
99.1(1.0) &
[3]\\

10& 
51$\;$295.029855(5) & 
27778&
92.5(4)&
[3]&
40&
52$\;$027.692627(9)&
30249&
99.2(7) &
[3]\\

11& 
51$\;$297.698476(8)&
27787& 
99.4(7)&  
[3]&
41&
52$\;$029.768118(8)&
30256&
95.8(7)  &
[3]\\

12&
51$\;$334.464970(12)&
27911&
93.6(1.1)&
[3]&
42&
52$\;$030.954185(12) &
30260&
100.0(1.1)  &
[3]\\

13&
51$\;$335.650973(6) &
27915 &
92.2(5) &
[3]&
43&
52$\;$032.733167(8)&
30266&
96.1(7)  &
[3]\\

14&
51$\;$337.133479(6)&
27920&
90.8(5) &
[3]&
44&
52$\;$076.615786(4)&
30414&
91.7(3) &
[3]\\

15&
51$\;$393.46935(4)&
28110&
92(4) &
[3]&
45&
52$\;$077.208801(7)&
30416&
92.2(6) &
[3]\\

16&
51$\;$396.13784(4)&
28119& 
87(3) &
[3]&
46&
52$\;$077.801847(6)&
30418&
95.3(5)  &
[3]\\

17&
51$\;$397.32378(3)&
28123&
81(3) &
[3]&
47&
52$\;$078.394837(7)  &
30420&
93.7(6)  &
[3]\\

18&
51$\;$466.112958(9)& 
28355&
91.5(8) &
[3]&
48&
52$\;$078.987831(8)& 
30422&
92.4(7)   &
[3]\\

19&
51$\;$467.29898(12)&
28359&
92.2(1.0) &
[3]&
49&
52$\;$131.469068(10) &
30599& 
86.8(9)   &
[3]\\

20&
51$\;$470.264016(9) & 
28369&
91.1(8)  &
[3]&
50&
52$\;$132.65509(2)&
30603&
87(2)   &
[3]\\

21&
51$\;$557.436333(6)&
28663&
89.8(5)  &
[3]&
51&
52$\;$133.24811(8) &
30605&
88(7) &
[3]\\

22&
51$\;$561.290937(6)&
28676&
93.7(5)  &
[3]&
52&
52$\;$136.50958(8)&
30616&
81(7) &
[3]\\

23&
51$\;$562.477008(6) &
28680& 
98.3(5)&
[3]&
53&
52$\;$159.34046(2)  &
30693&
83.9(1.4)  &
[3]\\

24&
51$\;$625.03951(2)&
28891 &
104(2)&
 [3] &
54&
57$\;$291.24010(2) &
48001 &
17(2)&
[3]\\

25& 
51$\;$677.817305(4)&
29069& 
101.3(4) &
 [3] &
55&
57$\;$294.20513(2)&
48011 &
16(2)&
[3]\\

26&
51$\;$681.671901(6)&
29082&
104.5(5)  &
 [3] &
56&
57$\;$499.682737(14)&
48704 &
13.7(1.2)&
[3]\\

27&
51$\;$682.857903(7)&
29086& 
103.2(6)  &
 [3] &
57&
 57$\;$792.03631(3)&
49690&
23(3)&
[3] \\

28&
51$\;$763.803676(5)&
29359& 
106.3(4) &
 [3] &
58&
57$\;$794.70484(5)&
49699&
22(4)&
[3] \\

29&
51$\;$764.989711(8)&
29363& 
107.7(7) &
 [3] &
59&
57$\;$795.89087 (5)&
49703&
23(4)&
[3] \\

30&
51$\;$768.84426(2)&
29376&
106.0(1.4)&
[4]\\

		\hline
	\end{tabular}

{\small \sc Note} 
\footnotesize ---  Epoch of reference 43$\;$058.72595 MJD, 
orbital period 7.11610872 hr, the associated errors are at 68\%
confidence levels; [1] \cite{Com_89}, [2]
\cite{Wachter_00}, [3] this work, [4] \cite{Ost_01} .
\end{table*} 
To estimate the eclipse arrival times, we folded the solar-system
barycentric corrected light curves using a trial time of reference and
orbital period, $T_{\rm fold}$ and $P_{\rm 0}$, respectively.  The value of the
adopted $T_{\rm fold}$ corresponds to a time close to the start time of
the corresponding observation.  The adopted value of $P_0$ is
7.11610872 hr corresponds to the value of the orbital period at
$T_0= 43\,058.72609$ MJD obtained by \cite{Ost_01} adopting quadratic
ephemeris.
 
We fitted the eclipse profiles with a simple model consisting of a
step-and-ramp function, where the count rates before, during, and
after the eclipse are constant and the intensity changes linearly
during the eclipse transitions. This model involves seven parameters:
the count rate before, during, and after the eclipse, called $C_1$,
$C_2$, and $C_3$, respectively; the phases of the start and stop times
of the ingress ($\phi_1$ and $\phi_2$), and, finally, the phases of
the start and stop times of the egress ($\phi_3$ and $\phi_4$).  We
show a typical eclipse of \mxbs in Fig. \ref{eclipse}. The eclipse was
observed during the RXTE/PCA observation P40050-04-16-00, the
superimposed red function is the step-and-ramp best-fitting function.
The phase corresponding to the eclipse arrival time $\phi_{\rm ecl}$ is
estimated as $\phi_{\rm ecl} = (\phi_2+\phi_3)/2$.  The corresponding
eclipse arrival time is given by
$T_{\rm ecl} = T_{\rm fold} + \phi_{\rm ecl} P_0 $. To be more conservative, we
scaled the error associated with $\phi_{\rm ecl}$ by the factor
$\sqrt{\chi^2_{\rm red}}$ to take into account values of $\chi^2_{\rm red}$ of
the best-fit model larger than one.  We show the obtained eclipse
arrival times in Barycentric Dynamical Time (TDB), in units of MJD, in
Tab. \ref{tab:table_delay}.

We used the 43 RXTE/PCA observations to estimate the average duration,
$\Delta T_{\rm ecl}$, $\Delta T_{\rm ing}$ and $\Delta T_{\rm egr}$ of the
eclipse, the ingress and the egress, respectively.  The values of
$\Delta T_{\rm ecl}$, $\Delta T_{\rm ing}$ and $\Delta T_{\rm egr}$ for each
eclipse are shown as function of the eclipse arrival times in
Fig. \ref{duration}.  We found that $\Delta T_{\rm ecl}$ is scattered
between 890 and 910 s. Fitting the values of eclipse duration with a
constant we obtained a $\chi^2(d.o.f.)$ of 561(42) and a best-fit
value of $\Delta T_{\rm ecl} = 899.1 \pm 0.6$ s at 68\% confidence level
(c.l.).  The ingress duration is scattered between 10 and 30 s while
the egress duration is scattered between 10 and 35 s. Fitting the
ingress duration values with a constant we obtained a $\chi^2(d.o.f.)$
of 457 (38) and a best-fit value of $\Delta T_{\rm ing} = 17.0 \pm 0.7$ s
at 68\% c. l., while, fitting the egress duration values we obtained a
$\chi^2(d.o.f.)$ of 560 (39) and a best-fit value of
$\Delta T_{\rm egr} = 16.7 \pm 0.9$ s at 68\% c. l.. The associated errors
were scaled by the factor $\sqrt{\chi^2_{\rm red}}$ to take a value of
$\chi^2_{\rm red}$ of the best-fit model larger than one into account. We
find that the average duration of the ingress and egress are
similar. We also show in Fig. \ref{duration} the occurrences of the
measured ingress, egress and duration using a bin of 3.1, 3.7 and 3.5
s, respectively.

\begin{figure}
	\includegraphics[width=\columnwidth]{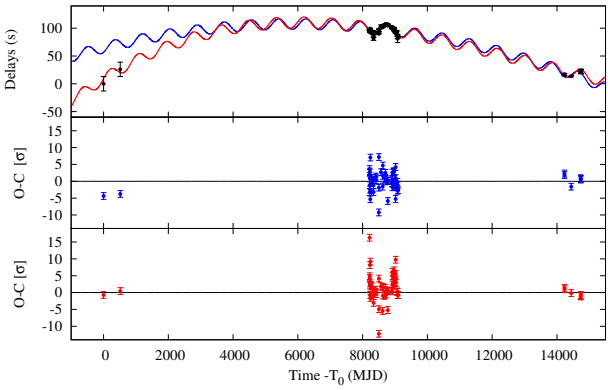}
        \caption{Top panel: delays with respect to the predicted eclipse arrival
          times, assuming as epoch of reference $T_0= 43\;058.72595$
          MJD and as orbital period $P_0= 7.11610872$ hr, plotted
          versus time. The  blue and red curves indicate the 
          best-fit functions corresponding to eqs. \ref{eq:sin} and
          \ref{eq:cub}, respectively. Middle panel: 
           residuals in units of $\sigma$ with respect to the
          blue curve.  Bottom panel: residuals
          in units of  $\sigma$  with respect to the red curve.} 
    \label{fig:residuals}
\end{figure}

 We calculated the delays with respect to $P_0= 7.11610872$ hr 
and to a reference epoch of 
$T_0=43\,058.72595$ MJD, corresponding to
 the first eclipse arrival time obtained by \cite{Com_89}. 
The inferred delays, in units of seconds, of the eclipse arrival times
with respect to a constant 
orbital period are reported in Tab. \ref{tab:table_delay}.
For each point we computed the corresponding cycle and the 
eclipse arrival time in days with respect to the adopted $T_0$. 
We show   the delays vs. time in Fig. \ref{fig:residuals} (top panel).
\begin{table} 
	\centering
	\caption{Best-fit values}
	\label{tab:best_fit}
	\begin{tabular}{l c   c c } 
		\hline
Parameter&
LQ&
LQS&
LQCS\\

		\hline
a  (s) &
$-109 \pm 38$ &
$ 65 \pm 20$ &
$ 9 \pm 29 $
\\
b (s d$^{-1}$) & 
$0.046 \pm 0.007$ &
$0.015 \pm 0.004 $ &
$ 0.037 \pm 0.009   $ 
 \\

c ($\times 10^{-6}$ s d$^{-2}$)&
$-2.6 \pm 0.3$&
$-1.2 \pm 0.2$&
$ -4.0 \pm 1.1$
\\
d ($\times 10^{-10}$ s d$^{-3}$)&
- &
 - &
$1.0 \pm 0.4$\\

A (s)&
- &
$9.6 \pm 0.6$&
$10.2 \pm 0.7$\\ 

 P$_{\rm mod}$ (d) &
-&
$843 \pm 7$&
$ 855 \pm 8$\\

 t$_\phi$ (d)&
-&
$137\pm75$ &
$ -7 \pm 82$\\

$\chi^2$(d.o.f.) &
4083(56)  &
512(53) &
455(52)\\
		\hline
	\end{tabular}
\end{table}

Initially we fitted the  delays with a quadratic function 
\begin{equation}
   y(t) = a+b t+ c t^2,
	\label{eq:quadratic}
\end{equation}
where $t$ is the time in days (MJD-43$\;$058.72595), $a=\Delta T_0$ is the
correction to $T_0$ in units of seconds, $b=\Delta P/P_0$ in units of
s d$^{-1}$ with $\Delta P$ the correction to the orbital period, and
finally, $c= 1/2 \;\dot{P}/P_0$ in units of s d$^{-2}$, with $\dot{P}$
representing the orbital period derivative.   The corresponding
  best-fitting 
parameters are shown in the LQ column of Tab. \ref{tab:best_fit}. 
With a $\chi^2 $ of 4083
for 56 d.o.f., we note that the quadratic function does not acceptably 
fit the data.
 Since the delays
seem to show a periodic modulation we fitted them using the
function
\begin{equation}
 y(t) = a+b t+ ct^2+A \sin\left[\frac{2 \pi}{P_{\rm {mod}}}(t-t_{\phi})\right],
	\label{eq:sin}
\end{equation}
where $A$ is the amplitude  in seconds of the sinusoidal function,
$P_{\rm {mod}}$ is the period of the sine function in days, and, finally,
$t_{\phi}$  represents  the time in days at which the sinusoidal function is
null.  A clear improvement is obtained with a value of $\chi^2({\rm {d.o.f.}})$ of
512 (53)  that translates to  a F-test probability
chance improvement of $7 \times 10^{-24}$.  The best-fit
function, indicated with a blue curve, and the corresponding residuals
are shown in the top and middle panels of Fig. \ref{fig:residuals}.
The best-fit values are shown in the third column of Tab.
\ref{tab:best_fit}.  The corresponding ephemeris (hereafter LQS) is
\begin{equation}
\label{linear_sin_eph}\begin{split}
T_{\rm ecl}(N) = {\rm MJD(TDB)}\; 43\;058.7267(2)  +
0.296504580(13) N -\\1.3(2) \times 10^{-12} N^2 +
A \sin\left[\frac{2 \pi}{N_{\rm {mod}}} N-\phi\right], 
\end{split}\end{equation}
where $N$ indicates the number of cycles, $N_{\rm {mod}}=P_{\rm {mod}}/P_0$ 
 and $\phi = 2 \pi t_{\phi}/P_{\rm {mod}} $. We  obtained an orbital period
 derivative  $\dot{P} = -8.5(1.2) \times 10^{-12}$ s s$^{-1}$, a  sinusoidal 
 modulation  characterised by a periodicity  $P_{\rm {mod}} = 2.31
 \pm 0.02$ yr
 and a semiamplitude  $A = 9.6 \pm 0.6$ s. 

It is evident that the LQS ephemeris does not predict the first two eclipse
arrival times. A possible explanation is that the orbital period
derivative is changing from 1976  up to now.  To take into account
 this fact, we added a cubic term to eq. \ref{eq:sin}, 
  defining the  new function
\begin{equation}
 y(t) = a+b t+ ct^2+dt^3+A \sin\left[\frac{2 \pi}{P_{\rm {mod}}}(t-t_{\phi})\right],
	\label{eq:cub}
\end{equation}  
where $d$ includes the presence of a derivative of  $\dot{P}$ with 
$d \simeq \ddot{P}/(6P)$.  With the latter model we obtain a value of  
$\chi^2({\rm {d.o.f.}})$ of 455 (52). By adding
the cubic term we find a F-test probability chance improvement of
0.014 indicating that the improvement of the fit is between two
and three $\sigma $ of confidence level.
The best-fit function, indicated
with a red curve, and the corresponding residuals are shown in the top
 and bottom panel of  Fig. \ref{fig:residuals}.  The best-fit
 parameters are shown in the fourth column of Tab. \ref{tab:best_fit}. The
corresponding ephemeris (hereafter LQCS) is
\begin{equation}
\label{linear_sin_eph}\begin{split}
T_{\rm ecl}(N) = {\rm MJD(TDB)}\; 43\;058.7261(3)  +
0.296504566(3) N -\\4.0(1.1) \times 10^{-12} N^2 +3.0(1.2)  \times 10^{-17} N^3+\\
A \sin\left[\frac{2 \pi}{N_{\rm {mod}}} N-\phi\right], 
\end{split}\end{equation}
 from which we inferred 
the orbital period derivative at  time $T_0=
43\;058.7261$ MJD to be 
 $\dot{P} = -2.7(7) \times 10^{-11}$ s s$^{-1}$ and the orbital period
 second derivative  $\ddot{P} = 2.4(9) \times 10^{-20}$ s s$^{-2}$. 
The  sinusoidal 
 modulation has a period of $P_{\rm {mod}} = 2.34 \pm 0.02$ yr and a semiamplitude 
of $A = 10.2 \pm 0.7$ s. 

\section{Discussion}

We  analysed the eclipse arrival times of \mxbs  with the main aim to estimate its
ephemeris.  Our baseline spans 40 years and covers the three outbursts
of the source observed from 1976. We  combined 51 eclipse
arrival times, corresponding to the outbursts occurred in 1999-2001 and
in 2015-2017,  with the  data  already reported in literature. 
 The   campaign of observations made with {\it Rossi-XTE}/PCA  during the
 1999-2001 outburst seems to indicate a possible periodic modulation
of 2.3 years; the delays associated with the six eclipse arrival times obtained during the most recent outburst agree with that periodic modulation.
We find that the LQS ephemeris  accounts for the eclipse arrival times
except for the two eclipses observed in 1976-1978. The addiction of a
cubic term (LQCS ephemeris) allows to  account for all the available
data, however the statistical improvement is less than three
sigma, suggesting that a larger baseline is needed to confirm the more complex
ephemerides. 
  In both cases, a sinusoidal modulation with a period between 840
  and 860 days is needed to obtain an acceptable fit of the eclipse arrival times.
In the following we  restrict our discussion to the LSQ ephemeris.

To estimate the eclipse arrival times we fitted the shape of the
  eclipse using a step-and-ramp function. We used the {\it RXTE}/PCA
  observations, covering 2.4 years during the second outburst of \mxb,
  to estimate the ingress/egress and eclipse durations. The obtained
  values are scattered, the mean values associated with the eclipse,
  ingress and egress are $\Delta T_{\rm ecl}= 899.1 \pm 0.6$ s,
  $\Delta T_{\rm ing}=17.0 \pm 0.7$ s and $\Delta T_{\rm egr}=16.7 \pm
  0.9$ s,  respectively.  We find that the
  ingress and egress durations are similar contrarily to what reported
  by \cite{Com_89}, that obtained an ingress and egress duration of
  $41 \pm 13$ s and $19 \pm 13$ s, respectively.  Our different
  results can be explained by the larger sample and the higher quality
   of our dataset.

  The ingress, egress, and eclipse durations show a jittered behaviour
  of the order of 15 s  similarly  to what    observed in \exos
  \citep{Wolff_02}.  \cite{Wolff_07} discussed the possibility that
  magnetic activity of the companion star generates extended coronal
  loops above the photosphere that could explain the amplitude of the
  observed jitter.  This scenario may be plausible  given the late
    K or early 
M type nature of the 0.3-0.4 M$_{\odot}$ companion star in \exo.
  Such
  stars can have magnetic activity if they rotate and if they have
  significant convective envelopes \cite[see][]{Wolff_07}.  The
  companion star in \mxbs is an early K type main-sequence star (see
  below), and hence it  likely has similar magnetic activity.  
  \cite{Ponti_17} showed that \axs has a different phenomenology.
  Although jitters are observed in the ingress and egress, the eclipse
  duration is nearly constant.  The authors suggested that
  the matter ejected from the accretion disc could reach the companion
  star with a ram pressure comparable to the pressure in the upper
  layers of the companion star (that is a K type main-sequence
  star). This interaction could displace the atmosphere of the
  companion star delaying both the ingress and the egress times.

\subsection{The masses of the binary system}
We can estimate the companion star  radius from the size of
its Roche lobe,  that  can be expressed by using the formula of
\cite{Pac_71} 
\begin{equation}
R_{\rm L_{2}}= 0.462 a \left(\frac{m_2}{m_1+m_2}\right)^{1/3},
\end{equation}
where $a$ is the orbital separation of the binary system and $m_{\rm 1}$ is
the neutron star mass in units of solar masses.  Combining the
previous equation with the third Kepler's law we find that
\begin{equation}
\label{rl2}
R_{\rm {L_{2}}} = 0.233 \;m_2^{1/3} P_{\rm h}^{2/3} {\rm R_{\odot}.}
\end{equation}
Assuming that
the companion star fills its Roche lobe then the radius of the
companion star $R_2$ coincides with $R_{\rm {L_{2}}}$.  
 To estimate the mass of the companion star we adopted the mass-radius
relation for a companion star in thermal equilibrium obtained
by studying the cataclysmic variable systems \citep[eq. 16
in][]{knigge_11} although LMXBs lie in a somewhat different
region of parameter space. We adopted the relation valid for large orbital periods
that is
\begin{equation}
\label{rknigge}
R_{\rm {{2}}} = 0.293 \pm 0.010 \left( \frac{M_2}{M_{\rm conv}}
\right)^{0.69 \pm 0.03} {\rm R_{\odot}},
\end{equation}
 where M$_{\rm conv}$ has a value of  $0.20 \pm 0.02$ M$_{\odot}$ and
 it is the mass of the convective region of the companion star. 
Combining the eqs. \ref{rl2} and \ref{rknigge} and taking into account that the
accuracy associated with the Roche lobe radius is 2\% we find that the
companion star has a mass of $0.9 \pm 0.3$ M$_{\odot}$ and a
radius of $0.84 \pm 0.10$ R$_{\odot}$.  Hereafter we will assume a
neutron star mass of $1.48 \pm 0.22$  M$_{\odot}$, this mass value was
estimated by \cite{Ozel_12} analysing the  mass distribution of neutron stars 
that have been recycled; the best value is the mean of the
distribution and the associated error is the corresponding dispersion.  

\subsection{The mass accretion rate and the mass transfer rate}

 Using RXTE/PCA data taken during the outburst in 1999,
  \cite{Gallo_08}  observed that the  flux of
\mxbs peaked at $\sim 1.0  \times 10^{-9}$ erg
s$^{-1}$ cm$^{-2}$ in the 2-25 keV energy range  during April 1999 ,
but it was between 4 $\times 10^{-10}$  and 6 $\times 10^{-10}$ erg s$^{-1}$ cm$^{-2}$ 
throughout the remainder of the outburst.
To infer a good estimation of  the flux in the 0.1-100 keV energy
band,   we adopted the broadband best-fit model of the persistent spectrum
obtained by \citealt{Ost_01}, from which we extrapolate an unabsorbed flux of
$1.0 \times 10^{-9}$ erg s$^{-1}$ cm$^{-2}$.

From the analysis of the type-I X-ray bursts the distance to \mxbs was
estimated to be $9 \pm2$ and $12 \pm 3$ kpc for a hydrogen-rich and
helium-rich companion star, respectively \citep[see][]{Gallo_08}.
We assume the average of the two values, $d=11 \pm4$ kpc, considering
that the source is accreting mixed H/He \citep[][]{Gallo_08}.

 To  convert the X-ray luminosity in mass accretion rate we used
  the relation
$L_x=\gamma \dot{M}_{\rm acc} c^2$, where $\gamma$ is the efficiency 
of the accretion and $c$ is the speed of the light. 
We take into account that the neutron star is rapidly spinning with a
frequency of 567 Hz \citep{Win_01} adopting 
the relation proposed by \cite{Sibga_00}  
\begin{equation}
\label{eff}\begin{split}
\gamma = 0.213 -0.153 f_{\rm{kHz}} +0.02 f_{\rm{kHz}}^2, 
\end{split}\end{equation}
where $f_{\rm{kHz}}$ is the spin frequency of the neutron star in
units of kHz.  The latter relation is valid assuming a gravitational
mass of the neutron star of 1.4 M$_{\odot}$ and the commonly 
adopted FPS equation of state for a neutron star. Using a spin
frequency of 567 Hz we find that $\gamma \simeq 0.132$.  Our
assumption implies that all of the released gravitational energy is
converted to X-ray emission and that negligible amount  of energy is carried away
by bulk outflows. This is confirmed by the spectral studies of the
source; in fact, the absorption lines associated with the presence of
\ion{Fe}{xxv} and \ion{Fe}{xxvi} ions are narrow suggesting that it is
not possible to associate to the source a superluminal jet
\citep[see][]{Sidoli_01}. Furthermore \cite{Diaz_16} suggested that
MXB 1659-298 could have a mild thermal wind but only static
atmospheres have been reported.

Using $\gamma \simeq 0.132$ we find
$ \dot{M}_{\rm acc}=(2.0 \pm 1.5) \times 10^{-9}$ M$_{\odot}$
yr$^{-1}$.  Considering a quiescence duration of almost 14.5 yr and a
mean outburst duration of 1.5 yr we find that the average value of the
observed mass accretion rate is
$|\left<\dot{M}_{\rm acc}\right>| \simeq \dot{M}_{\rm acc} 1.5/16 =
(1.9\pm 1.4) \times 10^{-10}$ M$_{\odot}$ yr$^{-1}$.

On the other hand, from theoretical considerations, we can estimate
the rate of mass that has to be transferred from the companion star in
order to explain the quadratic term of the LQS ephemeris interpreted
as the orbital period derivative of the system.  From the long-term
orbital evolution we can estimate the mass transfer rate
$\dot{M}_{\rm 2}$ using the eq.  4 in \cite{Burderi_10}
\begin{equation}
\label{mdot_in}\begin{split}
\dot{m}_{-8}=35(3n-1)^{-1}m_2\left(\frac{\dot{P}_{-10}}{P_{\rm 5h}}\right),
\end{split}\end{equation}
where $\dot{m}_{-8}$ is the mass transfer rate ${\dot{M}_2}$ in units
of $10^{-8}$ M$_{\odot}$ yr$^{-1}$, $n$ is the mass-radius index of
the companion star, ${m_2}$ is the companion star mass in units of
solar masses, $\dot{P}_{\rm -10}$ is the orbital period derivative in
units of $10^{-10}$ s s$^{-1}$ and $P_{\rm 5h}$ is the orbital period
in units of 5 hr. This is derived combining the third Kepler law with
the contact condition, that is
$\dot{R}_{\rm L2}/R_{\rm L2} = \dot{R}_2/R_2$ (where
$\dot{R}_{\rm L2}$ is the Roche Lobe radius of the secondary and $R_2$
is the radius of the secondary). Adopting $ {n = 0.69 \pm 0.03}$,
$ {m_2 = 0.9 \pm 0.3}$, ${\dot{P} = -8.5(1.2) \times 10^{-12}}$ s
s$^{-1}$ and ${P=7.1161099(3)}$ hr, we find that the mass transfer
rate implied by the observed orbital period derivative is
${\dot{M}_2=-(1.8 \pm 0.7) \times 10^{-8}}$ M$_{\odot}$ yr$^{-1}$,
that is almost two orders of magnitude higher that the observed
averaged mass accretion rate. This means that in order to explain the
observed orbital period change rate we have to invoke a highly not
conservative mass transfer for this system.

The above described scenario assumes a mass transfer rate of
${\dot{M}_2 = -(1.8 \pm 0.7) \times 10^{-8}}$ M$_{\odot}$ yr$^{-1}$
 and a companion star mass of $0.9 \pm 0.3$
M$_{\odot}$  in thermal equilibrium. The time scale associated with the mass transfer rate,
${\tau_{\rm \dot{M}}= M_2/|\dot{M}_2|}$, is $(5 \pm 3) \times 10^{7}$ yr.  The
companion star is in thermal equilibrium if ${\tau_{\rm \dot{M}}}$ is longer
than the thermal time scale ${\tau_{\rm KH}= GM_2^2/(R_2L_2)}$ of the
companion star \citep{Pac_71}.  To estimate the thermal timescale we
need to  infer the luminosity ${L_2}$ of the companion star. For a star
close to the lower main sequence it holds the relation
${L_2/L_{\rm \odot}=(M_2/M_\odot)^{4}}$ \citep[see][]{Salaris_05}.
 For a companion
star mass of $0.9 \pm 0.3$ M$_{\odot}$ we obtain that
${\tau_{\rm KH} = (5 \pm 3) \times 10^7}$ yr which is 
comparable with $\tau_{\rm \dot{M}} $,  for this reason we cannot
  exclude the the companion star is less massive of $0.9 \pm 0.3$ M$_{\odot}$.

\subsection{The prediction of the orbital period derivative for a
  highly non conservative mass transfer}
We can define a parameter $\beta$ in the following way, 
${ -\dot{M}_1=\beta \dot{M}_2}$, where 
${\dot{M}_1 =|\left<\dot{M}_{\rm acc}\right>| } $  is the mass
accretion rate. Hence $\beta=1$ in a conservative mass
transfer scenario and $\beta<1$ in a non conservative mass transfer
scenario. Comparing the observed averaged mass accretion rate with
the mass transfer rate implied by the observed orbital period derivative,   
we obtain $\rm{\beta= 0.011 \pm 0.009}$, suggesting that only $\sim 1\%$
of the mass transferred from the companion star is indeed accreted onto 
the neutron star.

According to the orbital evolution theory, orbital period changes are
expected to be driven by mass transfer from the companion to the
compact object, by emission of gravitational waves (GR) and/or by
magnetic braking (MB).  For orbital periods larger than two hours the
effects of MB dominate the orbital evolution of the binary system.
Following \cite{Verbunt_81}, \cite{Verbunt_93} and \cite{Tauris_01}
the torque associated with MB can be parametrised as
\begin{equation}
\label{mdot_3}\begin{split}
T_{\rm MB} =  8.4 (k^2)_{0.1} f^{-2} m_1^{-1} P_{\rm 2h}^2 q^{1/3}(1+q)^{2/3},
\end{split}\end{equation}
where ${f}$ is  a dimensionless parameter for which a value of either  0.79 
\citep{Skumanich_72} or 1.78 \citep{Smith_79} has been assumed, 
${ k=0.323}$ is the gyration radius for a  star   
with  mass of 0.8 M$_{\odot}$ \citep{Claret_90}, ${P_{\rm 2h}}$ is the
orbital period in units of two hours, ${q}$ is the mass ratio ${M_2/M_1}$
and, finally, ${m_1}$ is the mass of the compact object in units of
solar masses. Because ${T_{\rm MB}}$ depends on ${f^{-2}}$ the effects of
the MB on the derivative of the angular
momentum of the binary system  will be larger
for $f=0.79$ than for $f=1.78$.

We can calculate the secular orbital period derivative expected from
the non-conservative secular evolution of the system using the
relation
\begin{equation}
\label{pdot_1}\begin{split}
  \dot{P}_{-12} = 1.37 q (1+q)^{-1/3} m_1^{5/3} P_{\rm 2h}^{-5/3}
\left[\frac{1/3-n}{2g(\alpha,\beta,q) -1/3+n}\right]\times\\
[1+T_{\rm MB}],
\end{split}\end{equation}
where 
\begin{equation}
\label{pdot_2}\begin{split}
 g(\alpha,\beta,q) = 1-\beta q- \frac{1-\beta}{1+q}\left(\frac{q}{3}+\alpha\right)
\end{split}\end{equation}
\citep[see ][]{Disalvo_08,Burderi_09,Burderi_10}, 
where  $  {\dot{P}_{\rm -12}}$ is the orbital period derivative in units
of $10^{-12}$ s s$^{-1}$ and ${\alpha}$ is a dimensionless
parameter that quantifies the specific angular momentum 
of the ejected matter in the case of a non-conservative
mass transfer scenario.
The specific
angular momentum, ${l_{\rm ej}}$, with which the transferred mass is lost
from the system can be written in units of the specific angular
momentum of the secondary, that is
${\alpha = l_{\rm ej}/( \Omega_{\rm orb} r_2^2) = l_{\rm ej} P (M_1 + M_2)^2/(2\pi
a^2M^2_1)}$, where ${r_2}$ is the distance of the secondary star from the
centre of mass of the system, ${a}$ is the orbital separation and ${P}$ is
the orbital period of the binary system. 
For a neutron star mass of $1.48 \pm 0.22$ M$_{\odot}$ we obtain an 
orbital period derivative of $-(6 \pm 3)  \times 10^{-12}$ s s$^{-1}$, 
 which is compatible within one $\sigma$ with the value 
$\dot{P}=-(8.5 \pm 1.2)  \times 10^{-12}$ s s$^{-1}$ inferred from the eclipse
arrival times.

A highly non-conservative mass transfer in this
source may be justified by the fact that \mxbs is a fast spinning
neutron star \citep{Win_01}.
During  the quiescent periods, if the region around the neutron star is
free from matter up to the light cylinder radius, the radiation
pressure of the rotating magnetic dipole, given by the Larmor formula,
may be able to eject from the system the matter transferred by the
companion star  at the inner Lagrangian point, according to
the mechanism termed {\it radio ejection} and described in detail in
\cite{Burderi_01}.  Once significant temporary reduction of the
  mass accretion rate occurs, the neutron star can emit as a
  magnetic-dipole rotator and the radiation pressure from the pulsar
  may be able to eject  the matter transferred from the companion 
  out of the
  system.  We note that the disc instability model \citep[see the
  review of ][]{Lasota_01} may have a role in triggering
  the {\it radio ejection} and  starting a non conservative mass
  transfer. The {\it radio ejection} has been invoked to explain the
high orbital period derivative observed in the accreting millisecond
pulsar (AMSP) SAX J1808.4-3658 \citep[see][]{Disalvo_08,Burderi_09},
and, more recently, for the AMSP SAX J1748.9-2021 for which a high
orbital period derivative is also observed \citep{Sanna_16}.  We
therefore suggest that a similar mechanism could be  also at work for \mxb.

The above described scenario assumes a mass transfer rate of
${\dot{M}_2 = -(1.8 \pm 0.7) \times 10^{-8}}$ M$_{\odot}$ yr$^{-1}$
 and a companion star mass of $0.9 \pm 0.3$
M$_{\odot}$. The time scale associated with the mass transfer rate,
${\tau_{\rm \dot{M}}= M_2/|\dot{M}_2|}$, is $(5.1 \pm 2.7) \times 10^{7}$ yr.  The
companion star is in thermal equilibrium if ${\tau_{\rm \dot{M}}}$ is longer
than the thermal time scale ${\tau_{\rm KH}= GM_2^2/(R_2L_2)}$ of the
companion star \citep{Pac_71}.  To estimate the thermal timescale we
need to  infer the luminosity ${L_2}$ of the companion star. For a star
close to the lower main sequence it holds the relation
${L_2/L_{\rm \odot}=(M_2/M_\odot)^{4}}$ \citep[see][]{Salaris_05}.
  Since the companion
star mass is $0.9 \pm 0.3$ M$_{\odot}$ we obtain that
${\tau_{\rm KH} = (5 \pm 3) \times 10^7}$ yr which is 
comparable with $\tau_{\rm \dot{M}} $.  Since the two timescales
  are comparable we cannot exclude that the companion star is out of the
  thermal-equilibrium; hence,  the   value  of $0.9 \pm 0.3$
  M$_\odot$ has to be considered an upper limit to the companion star mass.

We note that for a mass of the companion star lower than $0.9 M_\odot$
the mass transfer rate would be also lower, because of the linear
dependence of ${\dot{M}_2}$ on ${m_2}$ in eq. \ref{mdot_in}.
Therefore, the minimum mass transfer rate is reached for a $m_2=0.35$
M$_{\odot}$.  This has to be considered as a lower limit to the mass
of the companion since below this mass the companion star is expected
to become fully convective and the magnetic braking switches off
\citep{Rappa_83}. For this limiting mass, the mass transfer rate is
$(7 \pm 3) \times 10^{-9}$ M$_{\odot}$ yr$^{-1}$.  However, a detailed
study of the evolution of this system is beyond the aims of this paper
and will be reported elsewhere. Here we note that the results
presented in this paper do not change significantly adopting a lower
mass for the companion star. Therefore, we will continue our
discussion assuming a companion star mass of $0.9 \pm 0.3$
M$_{\odot}$, keeping in mind that lower masses for the companion star
are also possible.

\subsubsection{The changes of the equivalent hydrogen column density
  ${N_{\rm H}}$ 
during the X-ray quiescence}

The mass ejected from the system can explain the  variable
  equivalent hydrogen column density ${N_{\rm H}}$ measured during the
  X-ray quiescence of the source.  \cite{Cack_08, Cack_13} measured
  two different ${N_{\rm H}}$ values of $(2.0 \pm 0.1) \times 10^{21}$
  cm$^{-2}$ and $(4.7 \pm 1.3) \times 10^{21}$ cm$^{-2} $,
  respectively, at different times, while the estimation of  ${N_{\rm
      H}}$ obtained by \cite{Dickey_90} is  $1.8 \times 10^{21}$
  cm$^{-2}$. 
 Here we suggest that the matter ejected from the system can
account for the additional absorption.
Most of the matter provided by the companion is ejected from the inner
Lagrangian point forming a circumbinary ring of matter around \mxb.
Because of the large inclination angle of the system it is possible
that the ejected matter interposes between the source and the
observer.  Local density inhomogeneities and/or changes in the mass
transfer rate could produce changes in the equivalent hydrogen column
as observed by \cite{Cack_08, Cack_13} during quiescence.

 We use the eq. 4 of
\cite{Iaria_13} to estimate the density of the ejected matter
\begin{equation}
\label{dens}\begin{split}
  n(r)  \simeq 6.9 \times 10^{11} (1-\beta) \zeta^{-1} \eta^{-1}
  \dot{m}_{\rm E}  (m_1+m_2)^{-1} P_{\rm h}^{-1} \left(\frac{r}{a}\right)^{-3/2},
\end{split}\end{equation}
where ${n(r)}$ is the density in units of cm$^{-3}$, ${r}$ is the distance
from the inner Lagrangian point, $\zeta$ is a parameter that takes
into account a non-spherical distribution of matter, $\eta$ a
parameter larger than 1, ${\dot{m}_{\rm E}}$ is the mass transfer rate in
units of Eddington mass accretion rate and ${a}$ is the orbital
separation of the binary system. Adopting a mass transfer rate
of ${ \lvert \dot{M}_2 \rvert = (1.8 \pm 0.7) \times 10^{-8}}$ M$_{\odot}$ yr$^{-1}$, an
orbital period of 7.116 hr, a companion star mass and a neutron star
mass of $0.9 \pm 0.3$ M$_{\odot}$ and $1.48 \pm 0.22$ M$_{\odot}$, respectively, we obtain
 ${ n(a) = (5 \pm 2)  \times 10^{10} (\zeta\eta)^{-1}} $ cm$^{-3}$.
Supposing a constant particle density along the line of sight, we can
determine the equivalent hydrogen column density ${N_{\rm H}}$ associated with
the neutral matter using ${N_{\rm H}=n(a) \times a}$,  where
${a = (1.74 \pm 0.10)  \times 10^{11}}$ cm. We find
${N_{\rm H} = (8 \pm 4)  \times 10^{21} (\zeta\eta)^{-1}} $ cm$^{-2}$. Since the
quantity $\zeta\eta$ is close to unity \citep[see][]{Iaria_13} we
find that the equivalent hydrogen column of the cold matter is
${N_{\rm H} = (8 \pm 4)  \times 10^{21} }$ cm$^{-2}$, that is of the same order of
magnitude of the changes observed during quiescence of the source and, furthermore,
it explains the discrepancy by a factor of two between the ${N_{\rm H}}$ values
measured by \cite{Cack_13} and \cite{Dickey_90}.

\subsubsection{The inclination angle of the source}

From our estimate of the duration of the eclipse ingress, that
  is
 $\Delta T_{\rm ing} \simeq 17$ s, we can estimate the size
of the  corona, $R_{\rm c}$, surrounding the central source using the relation
\begin{equation}
\frac{2 \pi }{P} a= \frac{2R_{\rm c} }{\Delta T_{\rm ing} },
\end{equation}
we find  $R_{\rm c} = (3.6 \pm 0.3) \times 10^8$ cm. 
Assuming a neutron star mass of $1.48 \pm 0.22$ M$_{\odot}$ and a companion star
mass of $0.9 \pm 0.3$ M$_{\odot}$ we infer that the Roche lobe radius,
$R_{{\rm L_{1}}}$ of the compact object is $5.8 \times 10^{10}$ cm.  The
radius of the accretion disc, ${R_{\rm d}}$, corresponds to the tidal radius
${R_{\rm T} \simeq 0.9 R_{\rm {L_{1}}}} $
\citep[see][eq. 5.122]{Frank_02}, hence the accretion disc radius is
$R_{\rm d} \simeq 5.3 \times 10^{10}$ cm. This result suggests that the
corona is much smaller than the accretion disk, and therefore it is
a relatively compact corona around the neutron star. 

\begin{figure}
	\includegraphics[width=\columnwidth]{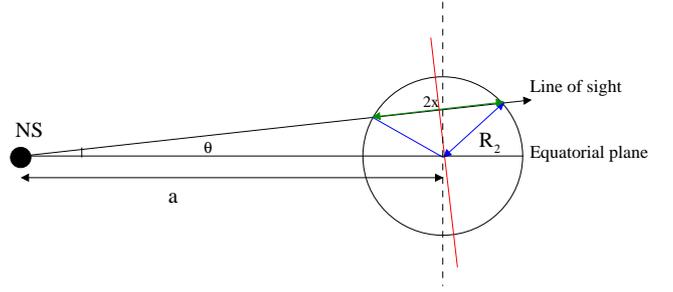}
  \caption{Schematic geometry of \mxbs not in scale.}
    \label{fig:geometry}
\end{figure}

Using our estimate of the eclipse duration we can also
estimate the inclination angle, $i=90^{\circ}-\theta$, of the system finding the
angle $\theta$  represented in Fig. \ref{fig:geometry}. 
Knowing that the eclipse duration is $\Delta T_{\rm ecl} \simeq 899.1$ s 
we can estimate the size of the occulted region $x$ as before using 
\begin{equation}
\frac{2 \pi }{P} a= \frac{2x }{\Delta T_{\rm ecl} }.
\end{equation}
We obtain $x =  (1.92 \pm 0.11) \times 10^{10}$ cm, where $2x$ is the
green segment shown in Fig. \ref{fig:geometry}. 
 The angle 
$\theta$,  representing the angle between the line of sight and the
equatorial plane of \mxb,  is obtained from 
$$
\tan \theta = \left[\frac{R_{2}^{2}-x^2}{a^2-(R_{2}^{2}-x^2)}\right]^{1/2}.
$$
We infer  $i =  72 \pm 3$ degrees. Our result is   compatible with the presence in
the light curve of the source of
dips and total eclipses that can be observed for  inclination angles
in the approximate range  $75^{\circ}$--80$^{\circ}$ \citep[see Fig. 5.10
in][]{Frank_02}. 
We note that for a companion star mass of 0.35 M$_{\odot}$  
 the inclination angle
of the system  is $76.0 \pm 0.7$ degree, that is marginally compatible
with the value obtained for a companion star mass  of $0.9 \pm 0.3$ M$_{\odot}$.

\cite{Sidoli_01} detected 
absorption lines associated with the presence of \ion{O}{viii},
\ion{Ne}{ix}, \ion{Fe}{xxv} and \ion{Fe}{xxvi} ions in the XMM spectrum of \mxb. The authors,
assuming an inclination angle of 80$^{\circ}$ inferred the distance 
 of the absorbing plasma from the central source, finding
$r_{{\rm Fe}} \la 2.4 \times 10^8$ cm, $r_{{\rm O}} \ga 3 \times 10^8$
cm and $r_{{\rm Ne}} \ga 9 \times 10^7$ cm, respectively.  Revisiting
the results obtained by \cite{Sidoli_01} for an inclination angle of
$72^{\circ}$ 
we find $r_{{\rm Fe}} \la 8 \times 10^8$ cm, $r_{{\rm O}} \ga 1 \times 10^9$
cm and $r_{{\rm Ne}} \ga 3 \times 10^8$.
Since we have estimated
a size of the corona of $R_{\rm c} \simeq 3.6 \times 10^8$ cm, we suggest
that the absorbing plasma is located  in the outer regions of the corona.

\subsection{The 2.31-yr periodic modulation: possible explanations}

Our ephemeris of  \mxbs  also includes a sinusoidal modulation with a
period of  $2.31 \pm 0.02$  yr. One possibility is that this
 periodic modulation observed in the delays may be produced by
the gravitational coupling of the orbit with changes in the shape of the
magnetically active companion star.  These changes are thought to be
the consequence of the torque applied by the magnetic activity of a
sub-surface magnetic field in the companion star with  the convective
envelope. The convective envelope induces a cyclic exchange of angular
momentum between the inner and outer regions of the companion star
causing a change in the gravitational quadrupole moment
\citep[see][]{Apple_92,Apple_94}. A similar mechanism has been
proposed for the eclipsing LMXBs \exos \citep{Wolff_09} and \axs
\citep{Ponti_17}.

The  inferred periodicity of 843 d and the
amplitude of 9.6 s  correspond in this case to an orbital period variation of 
$\Delta P/P = (8.3 \pm 0.5) \times 10^{-7}$.  We estimate that the
transfer of angular momentum needed to produce an orbital period
change $\Delta P$ is $\Delta J \simeq 3.8 \times 10^{46}$ g cm$^2$
s$^{-1}$ \citep[see][eq. 27]{Apple_92}.  The asynchronism of the
  companion, quantified through the quantity
$\Delta \Omega/\Omega$, is $3.7 \times 10^{-4}$, where $\Omega$ is the
orbital angular velocity of the binary system and $\Delta \Omega$ is
the variation of the orbital angular velocity needed to produce
$\Delta P$ \citep[see][eq. 3]{Apple_94}.  The variable part of the
luminosity of the companion star  required to power the
  gravitational
 quadrupole changes is
$\Delta L \simeq 1.5   \times 10^{32}$ erg s$^{-1}$. Considering that
$L_{\rm 2}/L_{\odot}=(M_{\rm 2}/M_{\odot})^{4}$ we obtain
$\Delta L/L_{\rm 2} = 0.06 \pm 0.10$,  in agreement
with the prediction of $\Delta L \simeq 0.1 L$ obtained for
magnetic active stars \citep[see][and references
therein]{Apple_92}. Our results suggest that a change in the magnetic
quadrupole of the companion star can produce the observed sinusoidal
modulation. The  energy required  to transfer the angular momentum
from the interior of the companion star to a thin shell, with a mass
of 10\% of M$_{\rm 2}$, at the surface (and viceversa) is furnished by 
 ten percentage of the thermonuclear energy produced by the companion star.
Furthermore, we obtain
that the mean sub-surface magnetic field $B$ of the companion star is
close to $1 \times 10^5$ G \citep[see ][eq. 23]{Apple_92}.

The origin of the sinusoidal modulation could also be explained by the
presence of a third body orbiting around the binary system, similarly
to what is found for the LMXB XB 1916-053 
\citep{Iaria_15}. Adopting the inclination angle of $72^{\circ}\!\!.1$
we find that the orbital separation between the centre of mass of
\mxbs and   the centre of mass of the triple system  is $a_{\rm x} \sin i= A\; c$, where $c$ is the
speed of light. Using the values in the third column of
Tab. \ref{tab:best_fit} we obtain that
$a_{\rm x} \sin i= (2.9 \pm 0.2) \times 10^{11}$ cm. Assuming a
non-eccentric and coplanar orbit of the third body  and that the
  companion star is in thermal equilibrium, the mass M$_{\rm 3}$ of
the third body is obtained from
\begin{equation}
\label{third_body}\begin{split}
  \frac{M_3 \sin i}{(M_3+M_{\rm bin})^{2/3}} =
  \left(\frac{4\pi^2}{G}\right)^{1/3} \frac{Ac}{P_{\rm mod}^{2/3}},
\end{split}\end{equation}
where M$_{\rm bin}$ is the mass of the binary system and $P_{\rm mod}$ is the
revolution period of the third body around the binary system
\citep[see e.g.][]{Bozzo_07}.   We obtain that the mass of the third
body is
$22 \pm 3$ M$_J$,  where M$_J$ indicates the Jovian mass;  
the distance of the third body from the centre of
mass of the triple system is $2.3 \pm 0.3$ AU. Releasing the constrain
of a co-planar orbit the mass of the third body is larger than 21 M$_J$. We have checked that the 
derived orbit of the third body is stable in the formalism 
by \cite{Kiseleva_94}. If this result will be confirmed, this will be
the first circumbinary Jovian planet spotted around a LMXB. In the
case of a no-coplanar orbit we find that the mass of the third body
should be larger than 21 M$_J$.

\section{Conclusions}

We have estimated 51 eclipse arrival times  for \mxbs when the source
was in outburst in 2000, 2001 and 2015 using {\it Rossi-XTE}, {\it
  XMM-Newton}, {\it NuSTAR} and {\it Swift/XRT} data. Combining these
times to the previous  ones reported in literature we obtain a baseline of
40 years, from 1976 to 2017, to constrain the ephemeris of the source.
The data are clustered in three temporal intervals covering six years
corresponding to the periods  when the source was in outburst.  In
the hypothesis that the companion star is in thermal equilibrium and
fills its Roche Lobe, we
estimate that the companion star mass is $0.9 \pm 0.3$ M$_{\odot}$,
in agreement with the possibility that the companion is an early K-type
main-sequence star as reported in literature.
 
Using RXTE/PCA data we have studied the profile of the total
eclipse observing jitters in the ingress/egress duration and eclipse
duration of about 10-15 s. The average values of the ingress, egress
and eclipse durations are $17.0 \pm 0.7$  s, $16.7 \pm 0.9$ s and
$899.1 \pm 0.6$ s, respectively. Using the average ingress and eclipse
duration
values  we find that the size of the corona surrounding the neutron star
is $R_{\rm c} = (3.6 \pm  0.3) \times 10^8$ cm and the inclination angle of the
 binary system is $ 72 \pm 3$ degree  assuming a companion star in thermal equilibrium.

We find that the eclipse arrival times are well  described by ephemeris
composed of a linear, a quadratic and a sinusoidal term.  We find  an
  orbital period derivative of $\dot{P} = -8.5(1.2) \times
10^{-12}$ s s$^{-1}$. We are able to explain the value of   $\dot{P}$ 
assuming   a highly non conservative mass transfer scenario.
We find that the mass transfer rate is $\dot{M}_2= -(1.8 \pm 0.7)
\times 10^{-8}$ M$_{\odot}$ yr$^{-1}$, and only $1\%$ of this mass is
observed to accrete onto the neutron star.  We also suggest that the
ejected matter  produces a local absorber with an equivalent  hydrogen
column density of $(8 \pm 4) \times 10^{21} $ cm$^{-2}$.  

The sinusoidal modulation has a period of $2.31 \pm 0.02$ yr and an
amplitude of $9.6 \pm 0.6$ s.  The 2.3-yr
periodic modulation of the orbital period can be explained either with
the presence of a gravitational quadrupole coupling of the orbit to 
a variable deformation of the magnetically active companion star or
with the presence of a third body orbiting around the binary
system.  In the second scenario we find that the mass of the third body
is larger than $21$ M$_J$.

Finally, we note that the first two eclipse arrival times, measured
during the outburst occurred in 1976-1978,  are marginally
 accounted for  the quadratic ephemeris. To fit them we adopted a more complex
ephemeris  taking into account the second derivative of the orbital
period. However, the statistical improvement is smaller than three
$\sigma$. A larger baseline is needed to confirm or discard  more
complex  ephemerides.

\section*{Acknowledgements}
This research has made use of data and/or software provided by the
High Energy Astrophysics Science Archive Research Center (HEASARC),
which is a service of the Astrophysics Science Division at NASA/GSFC
and the High Energy Astrophysics Division of the Smithsonian
Astrophysical Observatory.
This research has made use of MAXI data provided by RIKEN, JAXA and
the MAXI team.We are grateful to the \emph{Swift} team, and especially Kim Page, for their
assistance and flexibility in the scheduling of our ToO  observations.
This work was  partially supported by the  Regione Autonoma della
Sardegna through POR-FSE Sardegna 2007-2013, L.R. 7/2007,
Progetti di Ricerca di Base e Orientata, Project N. CRP-60529.
We also acknowledge a financial contribution from
the agreement ASI-INAF I/037/12/0. AR and AS gratefully acknowledge
the Sardinia Regional Government for its financial support
(P.O.R. Sardegna F.S.E. Operational Programme of the Autonomous
Region of Sardinia, European Social Fund 2007-2013 - Axis IV Human
Resources, Objective l.3, Line of Activity l.3.1.).
We also acknowledge fruitful discussions with the international team
on "The disk magnetosphere interaction around transitional ms pulsars"
supported by ISSI (International Space Science Institute), Bern".




\bibliographystyle{mnras}
\bibliography{biblio} 

\begin{thebibliography}{}
\makeatletter
\relax
\def\mn@urlcharsother{\let\do\@makeother \do\$\do\&\do\#\do\^\do\_\do\%\do\~}
\def\mn@doi{\begingroup\mn@urlcharsother \@ifnextchar [ {\mn@doi@}
  {\mn@doi@[]}}
\def\mn@doi@[#1]#2{\def\@tempa{#1}\ifx\@tempa\@empty \href
  {http://dx.doi.org/#2} {doi:#2}\else \href {http://dx.doi.org/#2} {#1}\fi
  \endgroup}
\def\mn@eprint#1#2{\mn@eprint@#1:#2::\@nil}
\def\mn@eprint@arXiv#1{\href {http://arxiv.org/abs/#1} {{\tt arXiv:#1}}}
\def\mn@eprint@dblp#1{\href {http://dblp.uni-trier.de/rec/bibtex/#1.xml}
  {dblp:#1}}
\def\mn@eprint@#1:#2:#3:#4\@nil{\def\@tempa {#1}\def\@tempb {#2}\def\@tempc
  {#3}\ifx \@tempc \@empty \let \@tempc \@tempb \let \@tempb \@tempa \fi \ifx
  \@tempb \@empty \def\@tempb {arXiv}\fi \@ifundefined
  {mn@eprint@\@tempb}{\@tempb:\@tempc}{\expandafter \expandafter \csname
  mn@eprint@\@tempb\endcsname \expandafter{\@tempc}}}

\bibitem[\protect\citeauthoryear{{Applegate}}{{Applegate}}{1992}]{Apple_92}
{Applegate} J.~H.,  1992, \mn@doi [\apj] {10.1086/170967}, \href
  {http://adsabs.harvard.edu/abs/1992ApJ...385..621A} {385, 621}

\bibitem[\protect\citeauthoryear{{Applegate} \& {Shaham}}{{Applegate} \&
  {Shaham}}{1994}]{Apple_94}
{Applegate} J.~H.,  {Shaham} J.,  1994, \mn@doi [\apj] {10.1086/174906}, \href
  {http://adsabs.harvard.edu/abs/1994ApJ...436..312A} {436, 312}

\bibitem[\protect\citeauthoryear{{Bahramian}, {Heinke}, {Wijnands}  \&
  {Degenaar}}{{Bahramian} et~al.}{2016}]{Bahram_16}
{Bahramian} A.,  {Heinke} C.~O.,  {Wijnands} R.,   {Degenaar} N.,  2016, The
  Astronomer's Telegram, \href
  {http://adsabs.harvard.edu/abs/2016ATel.8699....1B} {8699}

\bibitem[\protect\citeauthoryear{{Bozzo} et~al.,}{{Bozzo}
  et~al.}{2007}]{Bozzo_07}
{Bozzo} E.,  et~al., 2007, \mn@doi [\aap] {10.1051/0004-6361:20078444}, \href
  {http://adsabs.harvard.edu/abs/2007A%26A...476..301B} {476, 301}

\bibitem[\protect\citeauthoryear{{Burderi} et~al.,}{{Burderi}
  et~al.}{2001}]{Burderi_01}
{Burderi} L.,  et~al., 2001, \mn@doi [\apjl] {10.1086/324220}, \href
  {http://adsabs.harvard.edu/abs/2001ApJ...560L..71B} {560, L71}

\bibitem[\protect\citeauthoryear{{Burderi}, {Riggio}, {di Salvo}, {Papitto},
  {Menna}, {D'A{\`i}}  \& {Iaria}}{{Burderi} et~al.}{2009}]{Burderi_09}
{Burderi} L.,  {Riggio} A.,  {di Salvo} T.,  {Papitto} A.,  {Menna} M.~T.,
  {D'A{\`i}} A.,   {Iaria} R.,  2009, \mn@doi [\aap]
  {10.1051/0004-6361/200811542}, \href
  {http://adsabs.harvard.edu/abs/2009A%26A...496L..17B} {496, L17}

\bibitem[\protect\citeauthoryear{{Burderi}, {Di Salvo}, {Riggio}, {Papitto},
  {Iaria}, {D'A{\`i}}  \& {Menna}}{{Burderi} et~al.}{2010}]{Burderi_10}
{Burderi} L.,  {Di Salvo} T.,  {Riggio} A.,  {Papitto} A.,  {Iaria} R.,
  {D'A{\`i}} A.,   {Menna} M.~T.,  2010, \mn@doi [\aap]
  {10.1051/0004-6361/200912881}, \href
  {http://adsabs.harvard.edu/abs/2010A%26A...515A..44B} {515, A44}

\bibitem[\protect\citeauthoryear{{Burrows} et~al.,}{{Burrows}
  et~al.}{2005}]{Burrows_05}
{Burrows} D.~N.,  et~al., 2005, \mn@doi [\ssr] {10.1007/s11214-005-5097-2},
  \href {http://adsabs.harvard.edu/abs/2005SSRv..120..165B} {120, 165}

\bibitem[\protect\citeauthoryear{{Cackett}, {Wijnands}, {Miller}, {Brown}  \&
  {Degenaar}}{{Cackett} et~al.}{2008}]{Cack_08}
{Cackett} E.~M.,  {Wijnands} R.,  {Miller} J.~M.,  {Brown} E.~F.,   {Degenaar}
  N.,  2008, \mn@doi [\apjl] {10.1086/593703}, \href
  {http://adsabs.harvard.edu/abs/2008ApJ...687L..87C} {687, L87}

\bibitem[\protect\citeauthoryear{{Cackett}, {Brown}, {Cumming}, {Degenaar},
  {Fridriksson}, {Homan}, {Miller}  \& {Wijnands}}{{Cackett}
  et~al.}{2013}]{Cack_13}
{Cackett} E.~M.,  {Brown} E.~F.,  {Cumming} A.,  {Degenaar} N.,  {Fridriksson}
  J.~K.,  {Homan} J.,  {Miller} J.~M.,   {Wijnands} R.,  2013, \mn@doi [\apj]
  {10.1088/0004-637X/774/2/131}, \href
  {http://adsabs.harvard.edu/abs/2013ApJ...774..131C} {774, 131}

\bibitem[\protect\citeauthoryear{{Chou}}{{Chou}}{2014}]{Chou_14}
{Chou} Y.,  2014, \mn@doi [Research in Astronomy and Astrophysics]
  {10.1088/1674-4527/14/11/001}, \href
  {http://adsabs.harvard.edu/abs/2014RAA....14.1367C} {14, 1367}

\bibitem[\protect\citeauthoryear{{Claret} \& {Gimenez}}{{Claret} \&
  {Gimenez}}{1990}]{Claret_90}
{Claret} A.,  {Gimenez} A.,  1990, \mn@doi [\apss] {10.1007/BF00640716}, \href
  {http://adsabs.harvard.edu/abs/1990Ap%26SS.169..215C} {169, 215}

\bibitem[\protect\citeauthoryear{{Cominsky} \& {Wood}}{{Cominsky} \&
  {Wood}}{1984}]{Com_84}
{Cominsky} L.~R.,  {Wood} K.~S.,  1984, \mn@doi [\apj] {10.1086/162361}, \href
  {http://adsabs.harvard.edu/abs/1984ApJ...283..765C} {283, 765}

\bibitem[\protect\citeauthoryear{{Cominsky} \& {Wood}}{{Cominsky} \&
  {Wood}}{1989}]{Com_89}
{Cominsky} L.~R.,  {Wood} K.~S.,  1989, \mn@doi [\apj] {10.1086/167117}, \href
  {http://adsabs.harvard.edu/abs/1989ApJ...337..485C} {337, 485}

\bibitem[\protect\citeauthoryear{{Cominsky}, {Ossmann}  \& {Lewin}}{{Cominsky}
  et~al.}{1983}]{Com_83}
{Cominsky} L.,  {Ossmann} W.,   {Lewin} W.~H.~G.,  1983, \mn@doi [\apj]
  {10.1086/161113}, \href {http://adsabs.harvard.edu/abs/1983ApJ...270..226C}
  {270, 226}

\bibitem[\protect\citeauthoryear{{Di Salvo}, {Burderi}, {Riggio}, {Papitto}  \&
  {Menna}}{{Di Salvo} et~al.}{2008}]{Disalvo_08}
{Di Salvo} T.,  {Burderi} L.,  {Riggio} A.,  {Papitto} A.,   {Menna} M.~T.,
  2008, \mn@doi [\mnras] {10.1111/j.1365-2966.2008.13709.x}, \href
  {http://adsabs.harvard.edu/abs/2008MNRAS.389.1851D} {389, 1851}

\bibitem[\protect\citeauthoryear{{D{\'{\i}}az Trigo} \& {Boirin}}{{D{\'{\i}}az
  Trigo} \& {Boirin}}{2016}]{Diaz_16}
{D{\'{\i}}az Trigo} M.,  {Boirin} L.,  2016, \mn@doi [Astronomische
  Nachrichten] {10.1002/asna.201612315}, \href
  {http://adsabs.harvard.edu/abs/2016AN....337..368D} {337, 368}

\bibitem[\protect\citeauthoryear{{D{\'{\i}}az Trigo}, {Parmar}, {Boirin},
  {M{\'e}ndez}  \& {Kaastra}}{{D{\'{\i}}az Trigo} et~al.}{2006}]{Diaz_06}
{D{\'{\i}}az Trigo} M.,  {Parmar} A.~N.,  {Boirin} L.,  {M{\'e}ndez} M.,
  {Kaastra} J.~S.,  2006, \mn@doi [\aap] {10.1051/0004-6361:20053586}, \href
  {http://adsabs.harvard.edu/abs/2006A%26A...445..179D} {445, 179}

\bibitem[\protect\citeauthoryear{{Dickey} \& {Lockman}}{{Dickey} \&
  {Lockman}}{1990}]{Dickey_90}
{Dickey} J.~M.,  {Lockman} F.~J.,  1990, \mn@doi [\araa]
  {10.1146/annurev.aa.28.090190.001243}, \href
  {http://adsabs.harvard.edu/abs/1990ARA%26A..28..215D} {28, 215}

\bibitem[\protect\citeauthoryear{{Filippenko}, {Leonard}, {Matheson}, {Li},
  {Moran}  \& {Riess}}{{Filippenko} et~al.}{1999}]{Filippenko_99}
{Filippenko} A.~V.,  {Leonard} D.~C.,  {Matheson} T.,  {Li} W.,  {Moran} E.~C.,
    {Riess} A.~G.,  1999, \mn@doi [\pasp] {10.1086/316413}, \href
  {http://adsabs.harvard.edu/abs/1999PASP..111..969F} {111, 969}

\bibitem[\protect\citeauthoryear{{Frank}, {King}  \& {Raine}}{{Frank}
  et~al.}{2002}]{Frank_02}
{Frank} J.,  {King} A.,   {Raine} D.~J.,  2002, {Accretion Power in
  Astrophysics: Third Edition}

\bibitem[\protect\citeauthoryear{{Galloway}, {Muno}, {Hartman}, {Psaltis}  \&
  {Chakrabarty}}{{Galloway} et~al.}{2008}]{Gallo_08}
{Galloway} D.~K.,  {Muno} M.~P.,  {Hartman} J.~M.,  {Psaltis} D.,
  {Chakrabarty} D.,  2008, \mn@doi [\apjs] {10.1086/592044}, \href
  {http://adsabs.harvard.edu/abs/2008ApJS..179..360G} {179, 360}

\bibitem[\protect\citeauthoryear{{Gehrels} et~al.,}{{Gehrels}
  et~al.}{2004}]{Gerels_04}
{Gehrels} N.,  et~al., 2004, \mn@doi [\apj] {10.1086/422091}, \href
  {http://adsabs.harvard.edu/abs/2004ApJ...611.1005G} {611, 1005}

\bibitem[\protect\citeauthoryear{{Harrison} et~al.,}{{Harrison}
  et~al.}{2013}]{Harrison_13}
{Harrison} F.~A.,  et~al., 2013, \mn@doi [\apj] {10.1088/0004-637X/770/2/103},
  \href {http://adsabs.harvard.edu/abs/2013ApJ...770..103H} {770, 103}

\bibitem[\protect\citeauthoryear{{Iaria}, {di Salvo}, {Burderi}, {D'A{\'{\i}}},
  {Papitto}, {Riggio}  \& {Robba}}{{Iaria} et~al.}{2011}]{Iaria_11}
{Iaria} R.,  {di Salvo} T.,  {Burderi} L.,  {D'A{\'{\i}}} A.,  {Papitto} A.,
  {Riggio} A.,   {Robba} N.~R.,  2011, \mn@doi [\aap]
  {10.1051/0004-6361/201117334}, \href
  {http://adsabs.harvard.edu/abs/2011A%26A...534A..85I} {534, A85}

\bibitem[\protect\citeauthoryear{{Iaria}, {Di Salvo}, {D'A{\`i}}, {Burderi},
  {Mineo}, {Riggio}, {Papitto}  \& {Robba}}{{Iaria} et~al.}{2013}]{Iaria_13}
{Iaria} R.,  {Di Salvo} T.,  {D'A{\`i}} A.,  {Burderi} L.,  {Mineo} T.,
  {Riggio} A.,  {Papitto} A.,   {Robba} N.~R.,  2013, \mn@doi [\aap]
  {10.1051/0004-6361/201015305}, \href
  {http://adsabs.harvard.edu/abs/2013A%26A...549A..33I} {549, A33}

\bibitem[\protect\citeauthoryear{{Iaria} et~al.,}{{Iaria}
  et~al.}{2015a}]{Iaria_15b}
{Iaria} R.,  et~al., 2015a, \mn@doi [\aap] {10.1051/0004-6361/201423402}, \href
  {http://adsabs.harvard.edu/abs/2015A%26A...577A..63I} {577, A63}

\bibitem[\protect\citeauthoryear{{Iaria} et~al.,}{{Iaria}
  et~al.}{2015b}]{Iaria_15}
{Iaria} R.,  et~al., 2015b, \mn@doi [\aap] {10.1051/0004-6361/201526500}, \href
  {http://adsabs.harvard.edu/abs/2015A%26A...582A..32I} {582, A32}

\bibitem[\protect\citeauthoryear{{Jahoda}, {Swank}, {Giles}, {Stark},
  {Strohmayer}, {Zhang}  \& {Morgan}}{{Jahoda} et~al.}{1996}]{Jahoda_96}
{Jahoda} K.,  {Swank} J.~H.,  {Giles} A.~B.,  {Stark} M.~J.,  {Strohmayer} T.,
  {Zhang} W.,   {Morgan} E.~H.,  1996, in {Siegmund} O.~H.,  {Gummin} M.~A.,
  eds,  \procspie Vol. 2808, EUV, X-Ray, and Gamma-Ray Instrumentation for
  Astronomy VII. pp 59--70, \mn@doi{10.1117/12.256034}

\bibitem[\protect\citeauthoryear{{Jansen} et~al.,}{{Jansen}
  et~al.}{2001}]{Jansen_01}
{Jansen} F.,  et~al., 2001, \mn@doi [\aap] {10.1051/0004-6361:20000036}, \href
  {http://adsabs.harvard.edu/abs/2001A%26A...365L...1J} {365, L1}

\bibitem[\protect\citeauthoryear{{Kiseleva}, {Eggleton}  \& {Orlov}}{{Kiseleva}
  et~al.}{1994}]{Kiseleva_94}
{Kiseleva} L.~G.,  {Eggleton} P.~P.,   {Orlov} V.~V.,  1994, \mn@doi [\mnras]
  {10.1093/mnras/270.4.936}, \href
  {http://adsabs.harvard.edu/abs/1994MNRAS.270..936K} {270, 936}

\bibitem[\protect\citeauthoryear{{Knigge}, {Baraffe}  \& {Patterson}}{{Knigge}
  et~al.}{2011}]{knigge_11}
{Knigge} C.,  {Baraffe} I.,   {Patterson} J.,  2011, \mn@doi [\apjs]
  {10.1088/0067-0049/194/2/28}, \href
  {http://adsabs.harvard.edu/abs/2011ApJS..194...28K} {194, 28}

\bibitem[\protect\citeauthoryear{{Lasota}}{{Lasota}}{2001}]{Lasota_01}
{Lasota} J.-P.,  2001, \mn@doi [\nar] {10.1016/S1387-6473(01)00112-9}, \href
  {http://adsabs.harvard.edu/abs/2001NewAR..45..449L} {45, 449}

\bibitem[\protect\citeauthoryear{{Levine}, {Bradt}, {Cui}, {Jernigan},
  {Morgan}, {Remillard}, {Shirey}  \& {Smith}}{{Levine}
  et~al.}{1996}]{Levine_96}
{Levine} A.~M.,  {Bradt} H.,  {Cui} W.,  {Jernigan} J.~G.,  {Morgan} E.~H.,
  {Remillard} R.,  {Shirey} R.~E.,   {Smith} D.~A.,  1996, \mn@doi [\apjl]
  {10.1086/310260}, \href {http://adsabs.harvard.edu/abs/1996ApJ...469L..33L}
  {469, L33}

\bibitem[\protect\citeauthoryear{{Lewin}, {Hoffman}, {Doty}  \&
  {Liller}}{{Lewin} et~al.}{1976}]{Lewin_76}
{Lewin} W.~H.~G.,  {Hoffman} J.~A.,  {Doty} J.,   {Liller} W.,  1976, \iaucirc,
  \href {http://adsabs.harvard.edu/abs/1976IAUC.2994....2L} {2994}

\bibitem[\protect\citeauthoryear{{Matsuoka} et~al.,}{{Matsuoka}
  et~al.}{2009}]{Matsuoka_09}
{Matsuoka} M.,  et~al., 2009, \mn@doi [\pasj] {10.1093/pasj/61.5.999}, \href
  {http://adsabs.harvard.edu/abs/2009PASJ...61..999M} {61, 999}

\bibitem[\protect\citeauthoryear{{Mihara} et~al.,}{{Mihara}
  et~al.}{2011}]{Mihara_11}
{Mihara} T.,  et~al., 2011, \mn@doi [\pasj] {10.1093/pasj/63.sp3.S623}, \href
  {http://adsabs.harvard.edu/abs/2011PASJ...63S.623M} {63, S623}

\bibitem[\protect\citeauthoryear{{Negoro} et~al.,}{{Negoro}
  et~al.}{2015}]{Negoro_15}
{Negoro} H.,  et~al., 2015, The Astronomer's Telegram, \href
  {http://adsabs.harvard.edu/abs/2015ATel.7943....1N} {7943}

\bibitem[\protect\citeauthoryear{{Oosterbroek}, {Parmar}, {Sidoli}, {in't Zand}
   \& {Heise}}{{Oosterbroek} et~al.}{2001}]{Ost_01}
{Oosterbroek} T.,  {Parmar} A.~N.,  {Sidoli} L.,  {in't Zand} J.~J.~M.,
  {Heise} J.,  2001, \mn@doi [\aap] {10.1051/0004-6361:20011006}, \href
  {http://adsabs.harvard.edu/abs/2001A%26A...376..532O} {376, 532}

\bibitem[\protect\citeauthoryear{{{\"O}zel}, {Psaltis}, {Narayan}  \& {Santos
  Villarreal}}{{{\"O}zel} et~al.}{2012}]{Ozel_12}
{{\"O}zel} F.,  {Psaltis} D.,  {Narayan} R.,   {Santos Villarreal} A.,  2012,
  \mn@doi [\apj] {10.1088/0004-637X/757/1/55}, \href
  {http://adsabs.harvard.edu/abs/2012ApJ...757...55O} {757, 55}

\bibitem[\protect\citeauthoryear{{Paczy{\'n}ski}}{{Paczy{\'n}ski}}{1971}]{Pac_71}
{Paczy{\'n}ski} B.,  1971, \mn@doi [\araa]
  {10.1146/annurev.aa.09.090171.001151}, \href
  {http://adsabs.harvard.edu/abs/1971ARA%26A...9..183P} {9, 183}

\bibitem[\protect\citeauthoryear{{Ponti}, {De}, {Mu{\~n}oz-Darias}, {Stella}
  \& {Nandra}}{{Ponti} et~al.}{2017}]{Ponti_17}
{Ponti} G.,  {De} K.,  {Mu{\~n}oz-Darias} T.,  {Stella} L.,   {Nandra} K.,
  2017, \mn@doi [\mnras] {10.1093/mnras/stw2317}, \href
  {http://adsabs.harvard.edu/abs/2017MNRAS.464..840P} {464, 840}

\bibitem[\protect\citeauthoryear{{Rappaport}, {Verbunt}  \& {Joss}}{{Rappaport}
  et~al.}{1983}]{Rappa_83}
{Rappaport} S.,  {Verbunt} F.,   {Joss} P.~C.,  1983, \mn@doi [\apj]
  {10.1086/161569}, \href {http://adsabs.harvard.edu/abs/1983ApJ...275..713R}
  {275, 713}

\bibitem[\protect\citeauthoryear{{Salaris} \& {Cassisi}}{{Salaris} \&
  {Cassisi}}{2005}]{Salaris_05}
{Salaris} M.,  {Cassisi} S.,  2005, {Evolution of Stars and Stellar
  Populations}

\bibitem[\protect\citeauthoryear{{Sanna} et~al.,}{{Sanna}
  et~al.}{2016}]{Sanna_16}
{Sanna} A.,  et~al., 2016, \mn@doi [\mnras] {10.1093/mnras/stw740}, \href
  {http://adsabs.harvard.edu/abs/2016MNRAS.459.1340S} {459, 1340}

\bibitem[\protect\citeauthoryear{{Sibgatullin} \& {Sunyaev}}{{Sibgatullin} \&
  {Sunyaev}}{2000}]{Sibga_00}
{Sibgatullin} N.~R.,  {Sunyaev} R.~A.,  2000, \mn@doi [Astronomy Letters]
  {10.1134/1.1323277}, \href
  {http://adsabs.harvard.edu/abs/2000AstL...26..699S} {26, 699}

\bibitem[\protect\citeauthoryear{{Sidoli}, {Oosterbroek}, {Parmar}, {Lumb}  \&
  {Erd}}{{Sidoli} et~al.}{2001}]{Sidoli_01}
{Sidoli} L.,  {Oosterbroek} T.,  {Parmar} A.~N.,  {Lumb} D.,   {Erd} C.,  2001,
  \mn@doi [\aap] {10.1051/0004-6361:20011322}, \href
  {http://adsabs.harvard.edu/abs/2001A%26A...379..540S} {379, 540}

\bibitem[\protect\citeauthoryear{{Skumanich}}{{Skumanich}}{1972}]{Skumanich_72}
{Skumanich} A.,  1972, \mn@doi [\apj] {10.1086/151310}, \href
  {http://adsabs.harvard.edu/abs/1972ApJ...171..565S} {171, 565}

\bibitem[\protect\citeauthoryear{{Smith}}{{Smith}}{1979}]{Smith_79}
{Smith} M.~A.,  1979, \mn@doi [\pasp] {10.1086/130579}, \href
  {http://adsabs.harvard.edu/abs/1979PASP...91..737S} {91, 737}

\bibitem[\protect\citeauthoryear{{Str{\"u}der} et~al.,}{{Str{\"u}der}
  et~al.}{2001}]{struder_01}
{Str{\"u}der} L.,  et~al., 2001, \mn@doi [\aap] {10.1051/0004-6361:20000066},
  \href {http://adsabs.harvard.edu/abs/2001A%26A...365L..18S} {365, L18}

\bibitem[\protect\citeauthoryear{{Tauris}}{{Tauris}}{2001}]{Tauris_01}
{Tauris} T.~M.,  2001, in {Podsiadlowski} P.,  {Rappaport} S.,  {King} A.~R.,
  {D'Antona} F.,   {Burderi} L.,  eds,  Astronomical Society of the Pacific
  Conference Series Vol. 229, Evolution of Binary and Multiple Star Systems.
  p.~145 (\mn@eprint {} {astro-ph/0012077})

\bibitem[\protect\citeauthoryear{{Verbunt}}{{Verbunt}}{1993}]{Verbunt_93}
{Verbunt} F.,  1993, \mn@doi [\araa] {10.1146/annurev.aa.31.090193.000521},
  \href {http://adsabs.harvard.edu/abs/1993ARA%26A..31...93V} {31, 93}

\bibitem[\protect\citeauthoryear{{Verbunt}}{{Verbunt}}{2001}]{Verbunt_01}
{Verbunt} F.,  2001, \mn@doi [\aap] {10.1051/0004-6361:20000469}, \href
  {http://adsabs.harvard.edu/abs/2001A%26A...368..137V} {368, 137}

\bibitem[\protect\citeauthoryear{{Verbunt} \& {Zwaan}}{{Verbunt} \&
  {Zwaan}}{1981}]{Verbunt_81}
{Verbunt} F.,  {Zwaan} C.,  1981, \aap, \href
  {http://adsabs.harvard.edu/abs/1981A%26A...100L...7V} {100, L7}

\bibitem[\protect\citeauthoryear{{Wachter}, {Smale}  \& {Bailyn}}{{Wachter}
  et~al.}{2000}]{Wachter_00}
{Wachter} S.,  {Smale} A.~P.,   {Bailyn} C.,  2000, \mn@doi [\apj]
  {10.1086/308754}, \href {http://adsabs.harvard.edu/abs/2000ApJ...534..367W}
  {534, 367}

\bibitem[\protect\citeauthoryear{Warner}{Warner}{1995}]{warner_95}
Warner B.,  1995, Cataclysmic Variable Stars.
Cambridge Astrophysics, Cambridge University Press

\bibitem[\protect\citeauthoryear{{Wijnands}, {Strohmayer}  \&
  {Franco}}{{Wijnands} et~al.}{2001}]{Win_01}
{Wijnands} R.,  {Strohmayer} T.,   {Franco} L.~M.,  2001, \mn@doi [\apjl]
  {10.1086/319128}, \href {http://adsabs.harvard.edu/abs/2001ApJ...549L..71W}
  {549, L71}

\bibitem[\protect\citeauthoryear{{Wijnands}, {Nowak}, {Miller}, {Homan},
  {Wachter}  \& {Lewin}}{{Wijnands} et~al.}{2003}]{Wij_03}
{Wijnands} R.,  {Nowak} M.,  {Miller} J.~M.,  {Homan} J.,  {Wachter} S.,
  {Lewin} W.~H.~G.,  2003, \mn@doi [\apj] {10.1086/377122}, \href
  {http://cdsads.u-strasbg.fr/abs/2003ApJ...594..952W} {594, 952}

\bibitem[\protect\citeauthoryear{{Wolff}, {Hertz}, {Wood}, {Ray}  \&
  {Bandyopadhyay}}{{Wolff} et~al.}{2002}]{Wolff_02}
{Wolff} M.~T.,  {Hertz} P.,  {Wood} K.~S.,  {Ray} P.~S.,   {Bandyopadhyay}
  R.~M.,  2002, \mn@doi [\apj] {10.1086/341264}, \href
  {http://adsabs.harvard.edu/abs/2002ApJ...575..384W} {575, 384}

\bibitem[\protect\citeauthoryear{{Wolff}, {Wood}  \& {Ray}}{{Wolff}
  et~al.}{2007}]{Wolff_07}
{Wolff} M.~T.,  {Wood} K.~S.,   {Ray} P.~S.,  2007, \mn@doi [\apjl]
  {10.1086/522833}, \href {http://adsabs.harvard.edu/abs/2007ApJ...668L.151W}
  {668, L151}

\bibitem[\protect\citeauthoryear{{Wolff}, {Ray}, {Wood}  \& {Hertz}}{{Wolff}
  et~al.}{2009}]{Wolff_09}
{Wolff} M.~T.,  {Ray} P.~S.,  {Wood} K.~S.,   {Hertz} P.~L.,  2009, \mn@doi
  [\apjs] {10.1088/0067-0049/183/1/156}, \href
  {http://adsabs.harvard.edu/abs/2009ApJS..183..156W} {183, 156}

\bibitem[\protect\citeauthoryear{{in 't Zand}, {Heise}, {Smith}, {Cocchi},
  {Natalucci}, {Celidonio}, {Augusteijn}  \& {Freyhammer}}{{in 't Zand}
  et~al.}{1999}]{Zand_99}
{in 't Zand} J.,  {Heise} J.,  {Smith} M.~J.~S.,  {Cocchi} M.,  {Natalucci} L.,
   {Celidonio} G.,  {Augusteijn} T.,   {Freyhammer} L.,  1999, \iaucirc, \href
  {http://adsabs.harvard.edu/abs/1999IAUC.7138....1I} {7138}

\makeatother
\end{thebibliography}





\bsp	
\label{lastpage}
\end{document}